\documentclass[twocolumn,tighten]{aastex63}

\usepackage{graphicx}
\usepackage{txfonts}
\usepackage{url} 
\usepackage{multirow}
\usepackage{times,epsfig,amssymb}
\bibpunct{(}{)}{;}{a}{}{,} 
\usepackage{floatrow}

\newcommand{\xmm}{\emph{XMM-Newton}}

\def\IASF{INAF/IASF-Milano, Via A. Corti 12, 20133 Milano, Italy}
\def\KIEL{  Christian-Albrechts-Universit\"at zu Kiel, Leibnizstr. 11, 24118 Kiel, Germany}
\def\MIT{Kavli Institute for Astrophysics and Space Research, Massachusetts Institute of Technology, Cambridge, Massachusetts, USA}
\def\INFN{INFN, Sezione di Pavia, Via A. Bassi 6, 2700 Pavia, Italy}
\def\IUS{Scuola Universitaria Superiore IUSS Pavia, piazza della Vittoria 15, 27100 Pavia, Italy}

\begin{document} 

\title{The origin of the unfocused XMM-Newton background, its variability and lessons learned for ATHENA}



\author{Fabio Gastaldello}
\affiliation{\IASF}
\author{Martino Marelli}
\affiliation{\IASF}
\author{Silvano Molendi}
\affiliation{\IASF}
\author{Iacopo Bartalucci}
\affiliation{\IASF}
\author{Patrick K\"uhl}
\affiliation{\KIEL}
\author{Catherine E. Grant}
\affiliation{\MIT}
\author{Simona Ghizzardi}
\affiliation{\IASF}        
\author{Mariachiara Rossetti}
\affiliation{\IASF}
\author{Andrea De Luca}
\affiliation{\IASF}
\affiliation{\INFN}
\author{Andrea Tiengo}
\affiliation{\IUS}
\affiliation{\IASF}
\affiliation{\INFN}

\correspondingauthor{Fabio~Gastaldello}
\email{fabio.gastaldello@inaf.it}

\begin{abstract}
We analyzed the unexposed to the sky (\emph{outFOV}) region of the MOS2 detector on board \xmm\ covering 15 years of data amounting to 255 Ms.
We show convincing evidence that the origin of the unfocused background in \xmm\ is due to energetic protons, electrons and hard X-ray photons.
Galactic Cosmic Rays are the main contributors as shown by the tight correlation (2.6 \% of total scatter) with 1 GeV protons data of the SOHO EPHIN detector. 
Tight correlations are found with a proxy of the Chandra background rate, revealing the common source of background for detectors in similar orbits, and with the data of the EPIC Radiation Monitor (ERM), only when excluding Solar Energetic Particles events (SEPs).

The entrance to the outer electron belts is associated to a sudden increase in the \emph{outFOV} MOS2 rate and a spectral change. These facts support the fact that MeV electrons can generate an unfocused background signal.

The correlation between MOS2 \emph{outFOV} data and the SOHO EPHIN data reveals a term constant in time and isotropic similar to the one found in the study of the pn data. The most plausible origin of this component is hard unfocused X-ray photons of the Cosmic X-ray Background (CXB) Compton-scattering in the detector as supported by the strength of the signal in the two detectors with different thicknesses.

Based on this physical understanding a particle radiation monitor on board ATHENA has been proposed and it is currently under study. It will be able to track different species with the necessary accuracy and precision to guarantee the challenging requirement of 2\% reproducibility of the background.
\end{abstract}  

%

\section{Introduction}

The instrumental background has always played an important role in X-ray 
missions, limiting the
exploitation of scientific data in particular of diffuse sources of low
surface brightness such as the intra-cluster medium of galaxy clusters 
\citep{Molendi:17}. This is painfully true for the data collected by the 
European Photon Imaging Camera (EPIC) instruments, the two MOS CCD cameras 
\citep{Turner.ea:01} and the pn CCD camera \citep{Struder.ea:01} on-board the 
ESA \xmm\ mission \citep{Jansen.ea:01}.

The EPIC instrumental background can be separated into particle and 
electronic noise components. 
The knowledge of these components has been growing throughout the entire 
\xmm\ lifetime thanks to the many efforts 
involved in collecting suitable blank sky fields to be used as template 
background by the \xmm\ users \citep{Read.ea:03,Carter.ea:07}, the analysis of 
the \xmm\ Guest Observer Facility leading to the \xmm\ Extended Source Analysis 
Software \citep{Kuntz.ea:08,Snowden.ea:08}, the efforts of the \xmm\ SOC\footnote{http://xmm2.esac.esa.int/docs/documents/GEN-TN-0014.pdf} and the 
contributions of various research teams, including our own 
\citep{De-Luca.ea:04,Gastaldello.ea:07*1,Leccardi.ea:08}.
A summary table of the EPIC instrumental background components is available 
at this link\footnote{www.star.le.ac.uk/~amr30/BG/BGTable.html} based upon
\citet{Read.ea:03}.

The properties (temporal behaviour, spectral and spatial distribution) of the 
signal generated by the interaction of particles with the detectors and with 
the surrounding structure depend on the energy of the primary particles 
themselves. 
Low energy particles (mainly protons with energies of a few tens of keV) 
accelerated in the Earth magnetosphere can reach the detectors as they are 
concentrated by the telescope mirrors. Their interactions with the 
CCDs generate events which are indistinguishable from valid X-ray photons 
and cannot be rejected on-board. Sudden increases of the count rate are 
observed due to these particles, dubbed ``soft proton flares". 
The time scale is extremely variable, ranging from hundreds of seconds 
to several hours, while the peak count rate can be more than three orders 
of magnitude higher than the quiescent one. The extreme time variability is 
the fingerprint of this background component, the 
Soft Proton (SP) component \citep[see][and references therein]{Fioretti.ea:16}.
This component, totally unexpected before the Chandra and \xmm\ launch, is now
a concern for future X-ray missions. The study of this component in the \xmm\
data and its connection with the Earth's magnetosphere is actively pursued 
\citep{Walsh.ea:14,Salvetti.ea:17,Ghizzardi.ea:17,Gastaldello.ea:17,Kronberg.ea:20,Kronberg.ea:21}.

High energy particles (E $>$ a few tens of MeV) generate a signal which is mostly 
discarded on board on the basis of an upper energy threshold and of a 
pattern analysis of the events \citep{Lumb.ea:02}.
The unrejected part of this signal represents an important component of the 
EPIC instrumental background. The time scale of its variability has always 
been qualitatively estimated to be much longer than the length of a typical 
\xmm\ observation and it has been typically dubbed the ``quiescent'' background
component (also called Non X-ray Background, NXB in the literature) as opposed to the flaring SP component. 
There are two ways to measure the quiescent background in the EPIC detectors: 1) through the analysis
of portions of the detector not exposed to the sky (\emph{outFOV}) where 
neither sky X-ray photons nor soft protons concentrated by the mirrors are collected there. This is the reason why this component is also dubbed 
``un-concentrated'' background component and we will use this terminology 
throughout the paper; 2) through the study of the observations with 
the filter wheel in closed position (FWC): in this configuration, a 1mm 
thick aluminum window prevents X-ray photons and soft protons from reaching 
the detector. The \emph{outFOV} regions offer the advantage 
of a measurement of the unfocused background simultaneous with the observation. FWC observations allow to check and eventually correct for spatial variations 
of the NXB spectrum across the detector.
EPIC MOS has been generally preferred for studies of this component because of the relatively small \emph{outFOV} pn detector area and for the 
higher percentage of Out of Time (OOT) events (6.3\% in Full Frame operation 
mode or 2.3\% in Extended Full Frame operation mode for the pn as 
opposed to 0.35\% for MOS).  
Moreover, soft protons contamination of the unexposed area of the pn detector due to 
a different camera geometry with respect to MOS has recently been documented \citep{Marelli.ea:21}.

Several attempts have been made to gain insight into the origin of the un-concentrated background of X-ray detectors. \cite{Hall.ea:08}, through detailed simulations, showed the bulk of the EPIC MOS and pn background can be due to secondary electrons produced by the interaction of Cosmic Ray protons with the structure surrounding the detectors. More recent work on the WFI \citep{Kienlin.ea18} and XIFU \citep{Lotti.ea:18} detectors to fly onboard the next ESA Large Mission Advanced Telescope for High Energy Astrophysics \citep[ATHENA,][]{Nandra.ea:13} lead to similar results.

In this paper we present a detailed characterization of the unfocused \xmm\ 
background through the analysis of a large amount of \emph{outFOV} MOS2 data. The motivation for using only the MOS2 dataset is to ensure a dataset as uniform as possible with time: we excluded the MOS1 camera from our analysis because of the loss of two peripheral CCDs on March 2005 and on December 2012.
Our goal is to identify the origin of the the EPIC MOS un-concentrated background. To reach it we will employ data from experiments sensitive to particles as well as from other X-ray detectors.

We will use the knowledge we gain to discuss possible improvements to the background characterization of detectors onboard ATHENA.

The paper is organized as follows: in Sect.\ref{sec:XMMdataset} we describe the preparation and analysis of the \xmm\ \emph{outFOV} MOS2 dataset. In Sect.\ref{sec:externaldata} we describe the datasets that we will cross-correlate with the \xmm\ data, in particular in Sect.\ref{sec:EPHINdata} the data from the EPHIN detector on board the SOHO satellite, in Sect.\ref{sec:Chandradata} the data from Chandra and in Sect.\ref{sec:RADMONdata} the data from the EPIC Radiation Monitor also on board \xmm. The qualitative similarity of the behavior of the
four data-sets as a function of time is impressive and in Sect.\ref{sec:corrsolarcycle} we quantitatively cross-correlate the four data-sets. 
In Sect.\ref{sec:electronbelts} we address the origin of the background enhancements at the end of the \xmm\ orbits. Results are discussed in Sect.\ref{sec:discussion} and summarized in Sect.\ref{sec:summary}. 



\begin{figure}[htbp]
\centering
\includegraphics[width=1.1\textwidth]{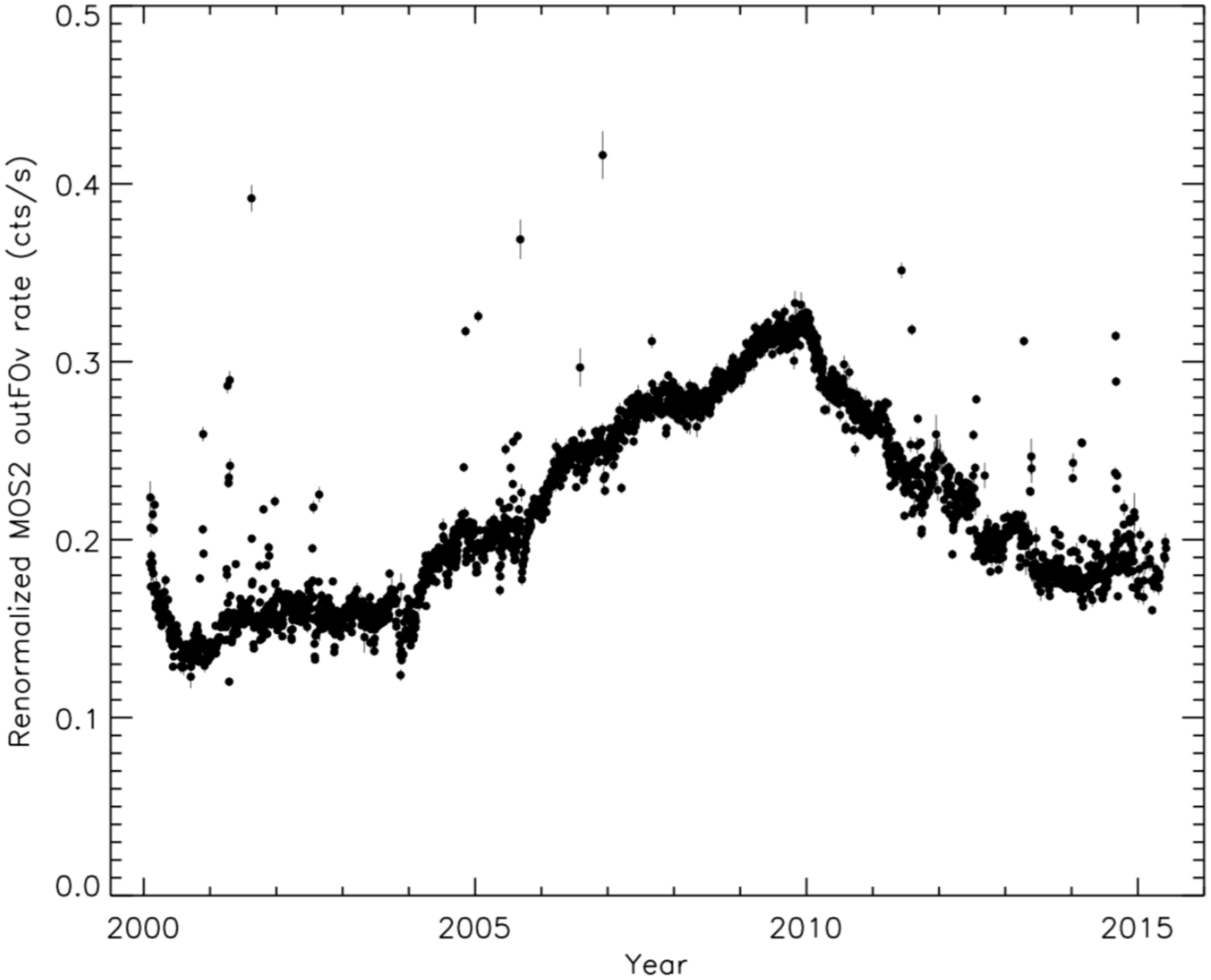}
\caption{MOS2 outFOV count rate (renormalized to the area of the full field of view) as a function of time. Each data point is the rate for a single \xmm\ revolution.}
\label{fig:1}%
\end{figure}

\begin{figure}[htbp]
\centering
\includegraphics[width=1.0\textwidth]{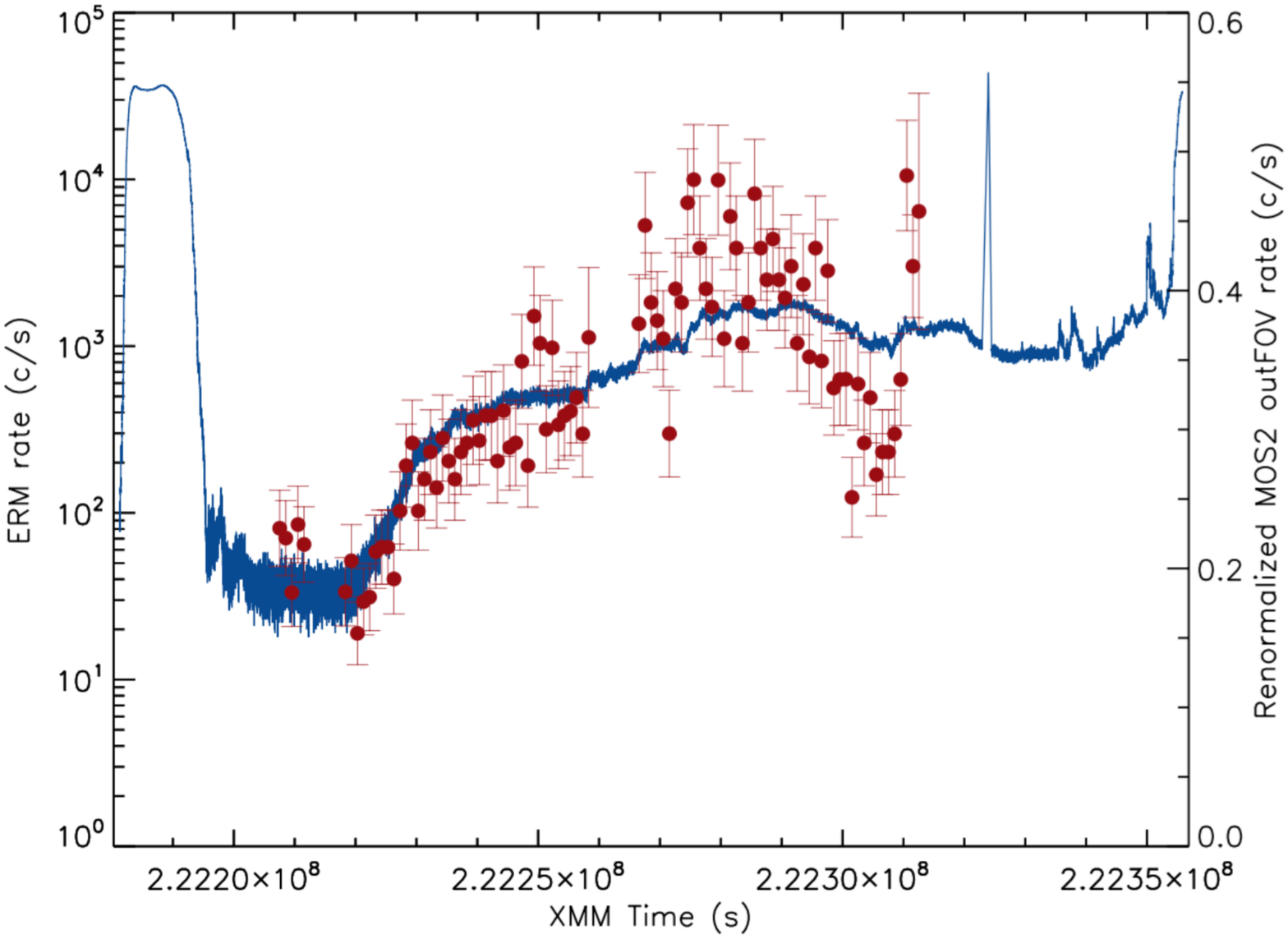}
\caption{MOS2 outFOV count rate of the data taken during revolution 935 (obsid 0201090201, 0202130301 and 0205340201) binned at 1ks and plotted as red circles with error bars and the corresponding ERM data in the HES0 channel as the blue solid line.}
\label{fig:2}%
\end{figure}

\begin{figure}[htbp]
\centering
\includegraphics[width=0.9\textwidth]{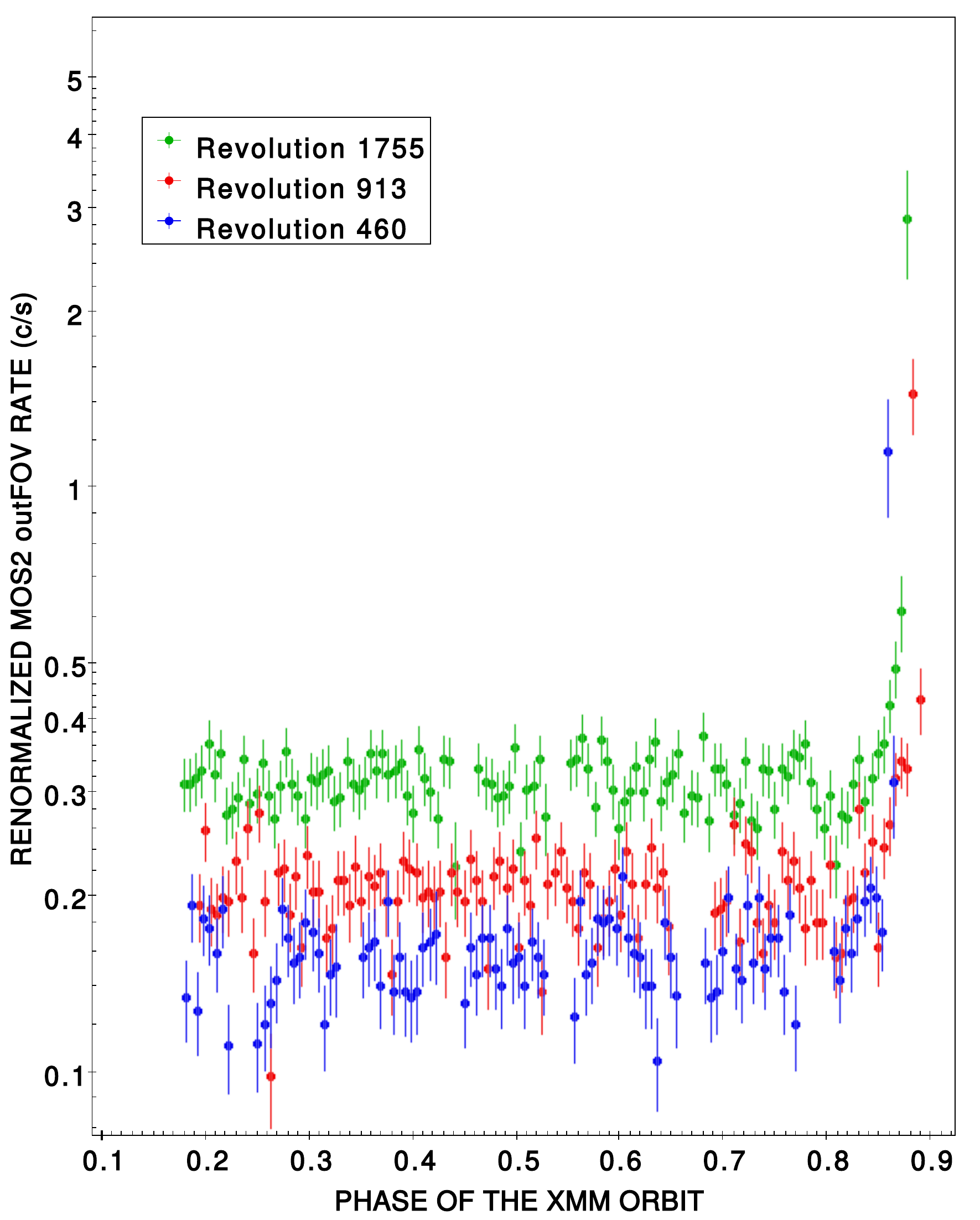}
\caption{Examples of MOS2 outFOV light curves as a function of the phase of the \xmm\ orbit showing that the variability is due to few ks at the end of the orbit.}
\label{fig:3}
\end{figure}

\section{The \xmm\ Dataset}
\label{sec:XMMdataset}
Our data set comprehends all the MOS2 exposures in the observations reported in the 3XMM source catalog distribution 6\footnote{http://xmmssc.irap.omp.eu/Catalogue/3XMM-DR6/3XMM\_DR6.html}: 10065 MOS2 exposures over the 9160 public \xmm\ EPIC observations made between 2000 February 3 and 2015 June 4. Differently from \citet{Marelli.ea:17}, we did not make any sub-selection based on the EXTraS project \citep{De-Luca.ea:21}, attitude variations, submode or celestial sources in the in-field-of-view since they are not relevant for the out-field-of-view (\emph{outFOV}) analysis.\\
We applied the same pattern, flags and energy filters, as well as the same data analysis, reported in \citet{Marelli.ea:17}. Light curves of the \emph{outFOV}  were extracted using the methods described in \citet{Marelli.ea:17}. We accumulated counts
in time bins of 1 ks and we further selected time bins
where the minimum number of counts accumulated in the \emph{outFOV}  area was 20 (the latter selection excluded 2.77 Ms of data) to ensure Gaussian errors. This is equivalent to excluding the time bins with a low exposed fraction of their duration as described in \citet{Salvetti.ea:17}. The dataset explored in this work covers 15 years of data and it amounts to 255 Ms of data.
The plot of all the available data binned in units of the \xmm\ revolution, which typically lasts 2 days and features an EPIC exposure time  of 120 ks, is shown in Fig.\ref{fig:1}. As we can see, some data points deviate strongly from the general trend followed by the bulk of the data: those are \xmm\ revolutions affected by Solar Energetic Particle (SEP) events. An example is reported in Fig.\ref{fig:2} where we show data from revolution 935 overplotted with the corresponding data from the \xmm\ radiation monitor in the HES0 channel (sensitive to protons in the 8-40 MeV range, see Sec \ref{sec:RADMONdata}).
To filter out \xmm\ revolutions during SEP events we used the SEP event list found on the ESA Solar Energetic Particle Environment Modelling (SEPEM) application server\footnote{http://www.sepem.eu/help/event\_ref.html}. That event list is updated only to 15-03-2013 and to include more recent SEP events we adopted the same parameters used for the generation of the reference list, i.e. proton flux above 0.001 cm$^{-2}$ s$^{-1}$ sr$^{-1}$ MeV$^{-1}$ in the energy channel 7.23-10.46 MeV.   
We therefore excluded 273 revolutions contaminated by SEP events ($\sim 10\%$ of the 2615 revolutions of our sample).
The total amount of time excluded corresponds to 20.2 Ms of data, 8\% of our total data and we are left with 235 Ms of data. 

We investigated if some variability is left at the timescale of the length of the typical observing time
during a single \xmm\ revolution. We tested the light-curves for each \xmm\ revolution for variability 
by calculating a $\chi^2$ test for the null hypothesis of a constant source count rate, flagging as variable
light curves with a probability $\leq 10^{-5}$ of being constant, as routinely applied in variability studies \citep[e.g. to asses variability in 3XMM sources,][]{Rosen.ea:16}. We found variability in 166 revolutions,
7\% of our SEP filtered sample of revolutions: for the great majority of the cases we found that this variability occurs in last few ks of the orbit when the \xmm\ satellite is entering the particle belts. 
As an example, in Fig.\ref{fig:3}, we show 3 light curves of revolutions during different years of the \xmm\ lifetime as a function of the phase of the \xmm\ orbit\footnote{We calculated the phase of the orbit by parsing the TSTART and TSTOP keywords of the orbit file available for each \xmm\ revolution at the \xmm\ webpage https://www.cosmos.esa.int/web/xmm-newton/radmon. They are the results of the \xmm\ Science Analysis Software (SAS) {\it orbit} task executed with a sample interval of 1 second.}. The origin of these enhancements will be discussed in Sect.\ref{sec:electronbelts}. 

\begin{figure}[htbp]
\includegraphics[width=0.9\textwidth]{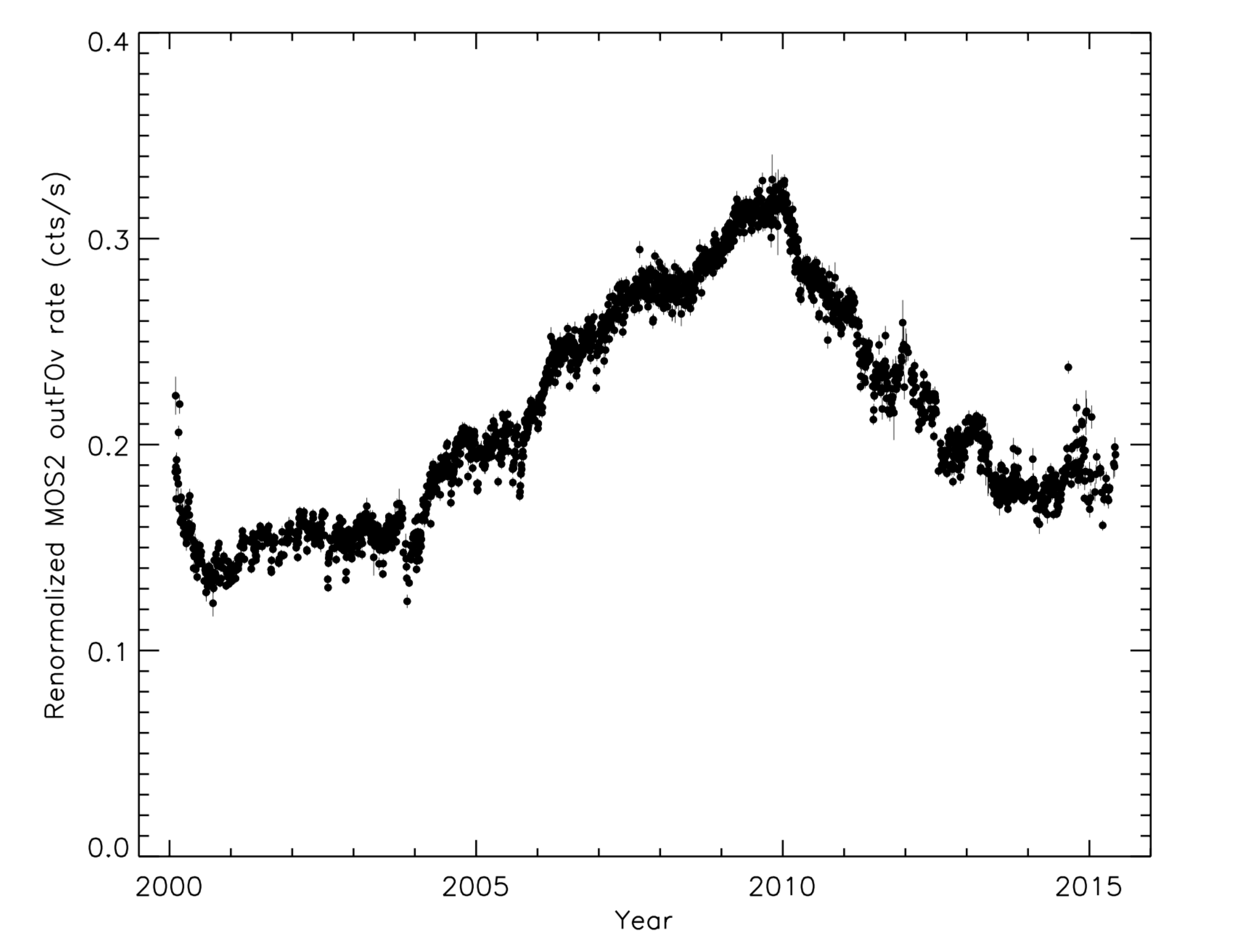}
\caption{Same as Fig.\ref{fig:1} filtered for data taken during SEP events and at the end of the revolution while entering the particle belts.}
\label{fig:4}%
\end{figure}

If we exclude the last periods of data acquisition of the various orbits by excluding data taken at phases of the orbit greater than 0.8, then the number of orbits exhibiting variability decreases to only 14. For these orbits the variability is in general due to a time bin with a low fractional exposure or, for some early data, when the entrance into the belt happens at smaller phases of the \xmm\ revolution (e.g. revolution 168 when the entrance into the belts happens at a phase of 0.7). If we exclude these data we are left with a data set of 2328 revolutions for 208 Ms of data. This will be our final cleaned sample used for the comparison with the external data sets detailed in the following sections and is shown in Fig.\ref{fig:4}. The overall reduction of scatter in the light curve has been almost totally achieved by the filtering of SEP events; the exclusion of the last phases of the \xmm\ orbits leads to a correction to the count-rate accumulated during each orbit of the order of 1\%.

The only exception to the use of this cleaned dataset will be for the discussion of Section \ref{sec:xmmerm} where we will revert to the use of the data not filtered by SEP events.

\section{External datasets}
\label{sec:externaldata}
\subsection{SOHO EPHIN data}
\label{sec:EPHINdata}

\begin{figure}[hb]
\includegraphics[width=0.9\textwidth]{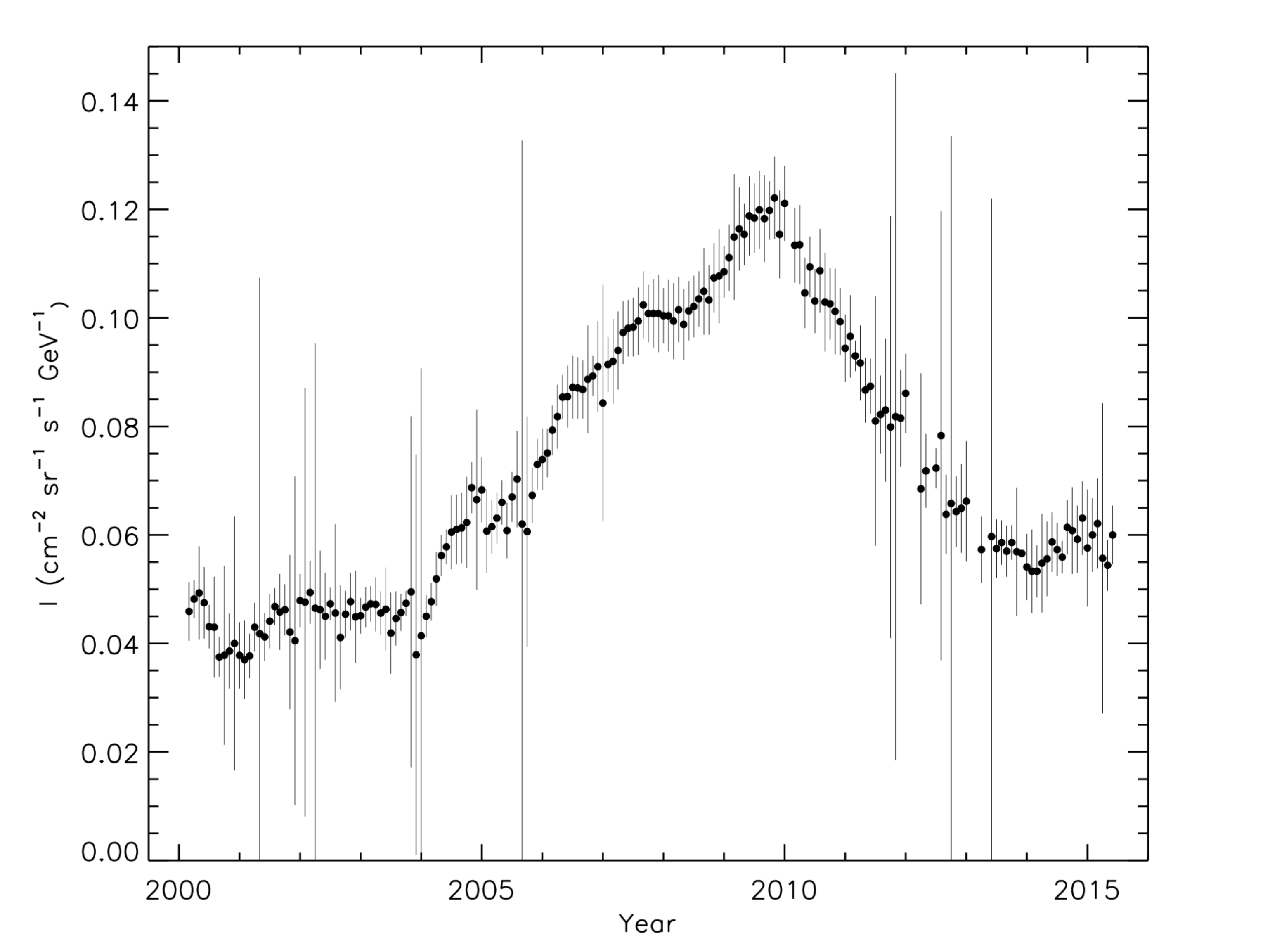}
\caption{SOHO EPHIN flux in the energy range centered at 1.04 GeV in the time period overlapping the available \xmm\ data presented in this work.}
\label{fig:5}%
\end{figure}

To access directly the relevant energy dependent flux of galactic cosmic rays (GCRs) we used measurements from EPHIN onboard SOHO. \citet{Kuehl.ea:16} have extended the energy range of the instrument to cover the energy range 250 MeV - 1.6 GeV, making available a large data set in terms of energy and time coverage (from 1996 to 2015) comparable to the \xmm\ \emph{outFOV} dataset. 
Here, we use monthly averaged fluxes measured between January 2000 and March 2015 and we show the data in the energy range centered on 1.04 GeV, see Fig.\ref{fig:5}. The systematic uncertainties of the fluxes are estimated to be below 20\% while, for most of the data points, the statistical uncertainty for GCRs is negligible with respect to the systematic one.

By comparing Figs.\ref{fig:4} and \ref{fig:5}, we can readily appreciate the  similarity  between the MOS2 \emph{outFOV} and EPHIN light curves. The quantitative correlation of the data will be discussed in Sect.\ref{sec:xmmephin}.

\subsection{Chandra rejected high energy data}
\label{sec:Chandradata}
\begin{figure}[hb]
\includegraphics[width=0.7\textwidth,angle=-90]{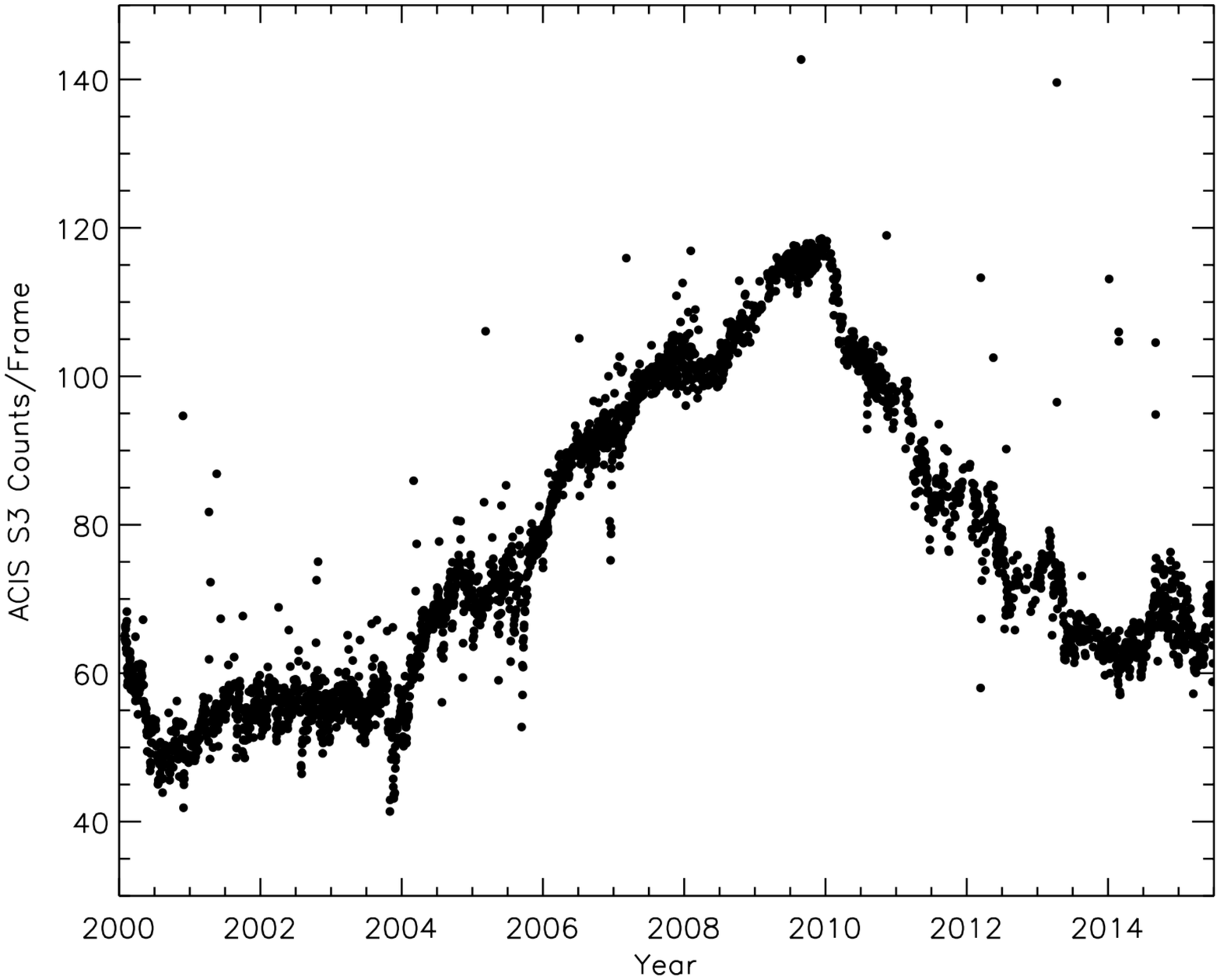}
\caption{Chandra high energy ($>$15 keV) count rate for the ACIS-S3 CCD as a function of time.  Each data point is the average rate for a single 3-10~ksec observation. The y-axis range has been clipped to a value of 150 counts/frame for clarity of the plot.}
\label{fig:6}%
\end{figure}

The Chandra X-ray Observatory is in a high-Earth orbit and ACIS, the Advanced CCD Imaging Spectrometer, uses CCDs, much like \xmm.  A proxy for the particle background on ACIS is the rate of events that exceed a high energy threshold, $\sim$15~keV, while the instrument is stowed and not viewing the sky. These events are identified and discarded on board, but the number in each frame is telemetered to the ground. 
The stowed position ensures that soft protons do not reach the detectors and therefore these data should be similar to MOS2 \emph{outFOV} data.
Figure~\ref{fig:6} shows the high energy reject rate for ACIS-S3, a back-illuminated CCD as a function of time \citep{Grant.ea:14}$\footnote{http://space.mit.edu/~cgrant/cti/cti120/bkg.pdf}$. Each bin represents a single observation: the mean (and median) exposure time is 8 ks.
The shape of the lightcurve of this ACIS background proxy 
mirrors those of the \xmm\ MOS2 \emph{outFOV} rate and of the EPHIN flux. The bulk of the outliers from the general trend can again be identified with SEP events which have been excluded from the analysis adopting the same methodology used for the MOS2 data explained in Sect.\ref{sec:XMMdataset}.

\begin{figure}[ht]
\includegraphics[width=1.0\textwidth]{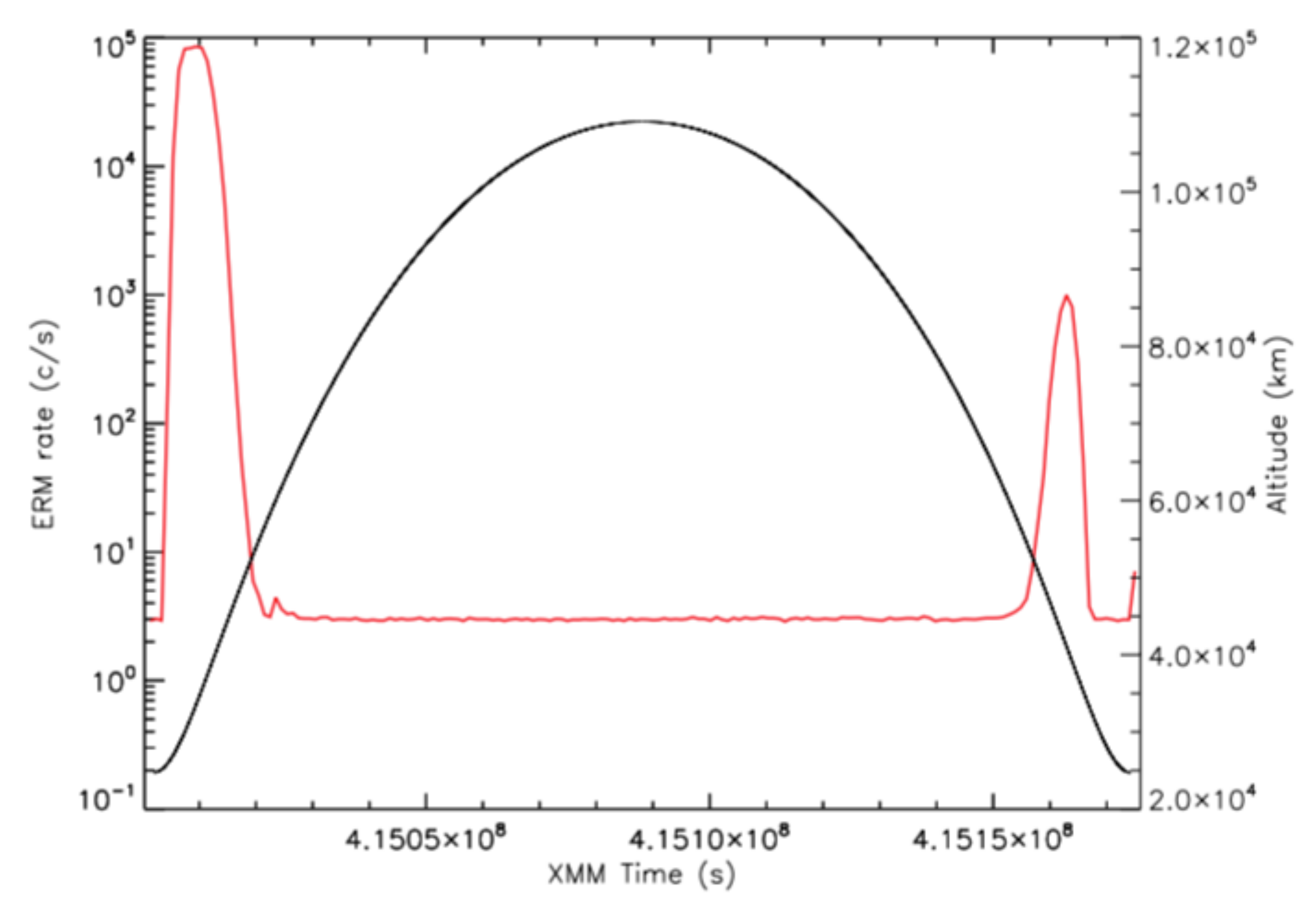}
\caption{ERM count rates in bins of 1ks for the HES0 channel (red line). Over-plotted with a black curve is the altitude of the \xmm\ orbit.} 
\label{fig:7}%
\end{figure}

\begin{figure}[ht]
\includegraphics[width=1.0\textwidth]{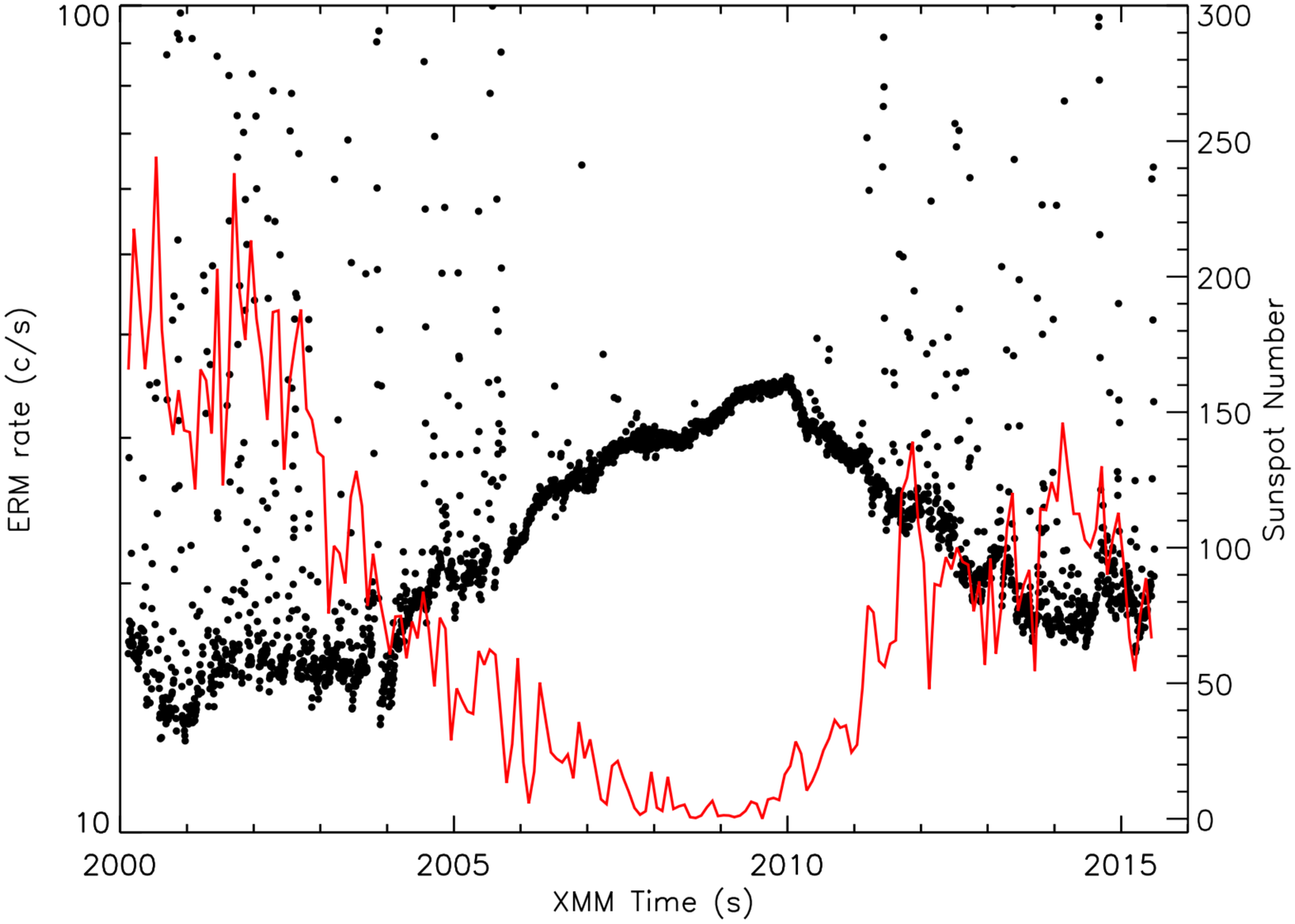}
\caption{The median count rate of the ERM HES0 in each \xmm\ orbit is shown as a function of time. The y-axis range has been clipped to a value of 100 cts/s and error bars have been omitted for clarity of the plot.
The number of sun spots is over-plotted with a red line.} 
\label{fig:8}%
\end{figure}

\begin{figure*}[hb]
\begin{center}
\resizebox{0.8\textwidth}{!}{
\includegraphics[]{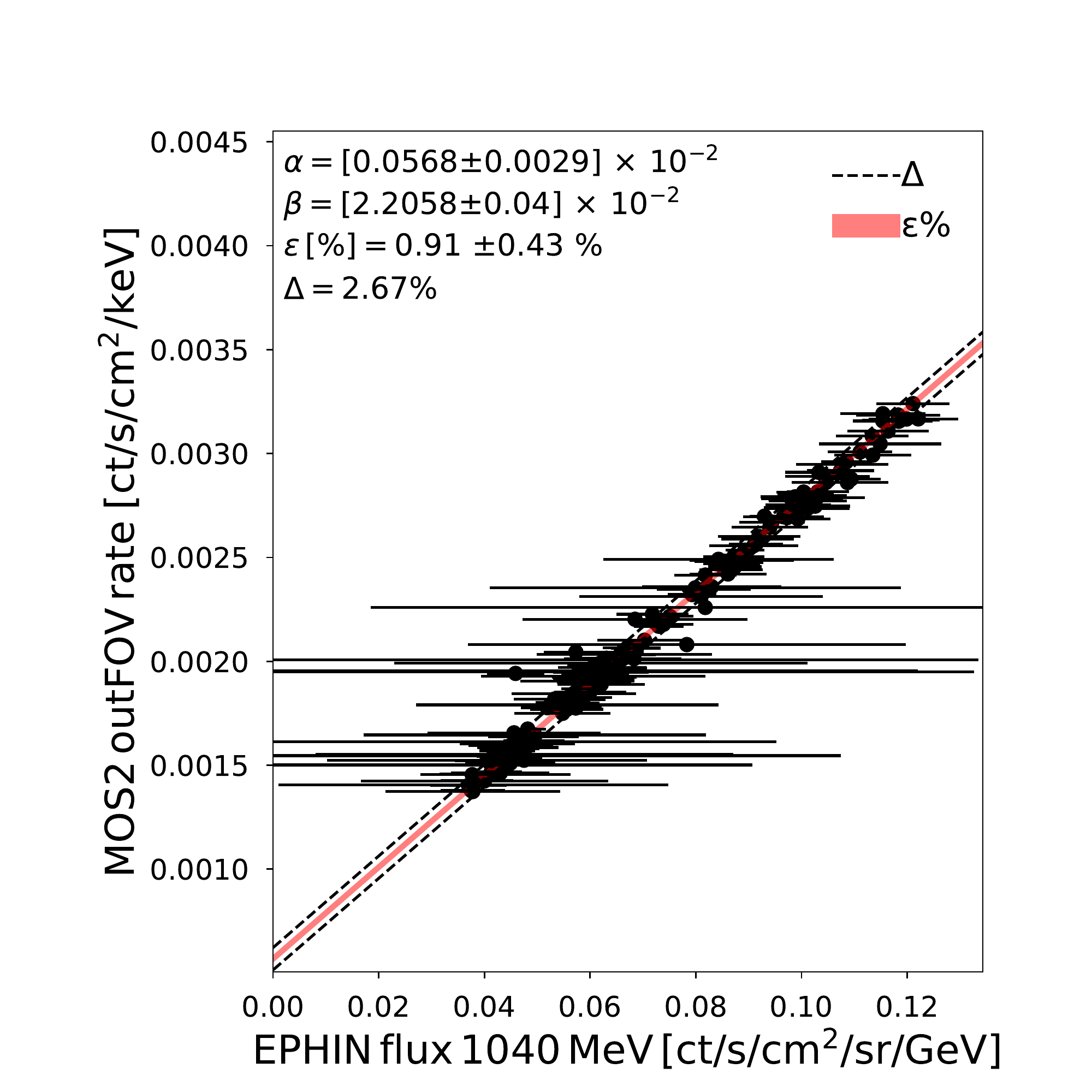}
\includegraphics[]{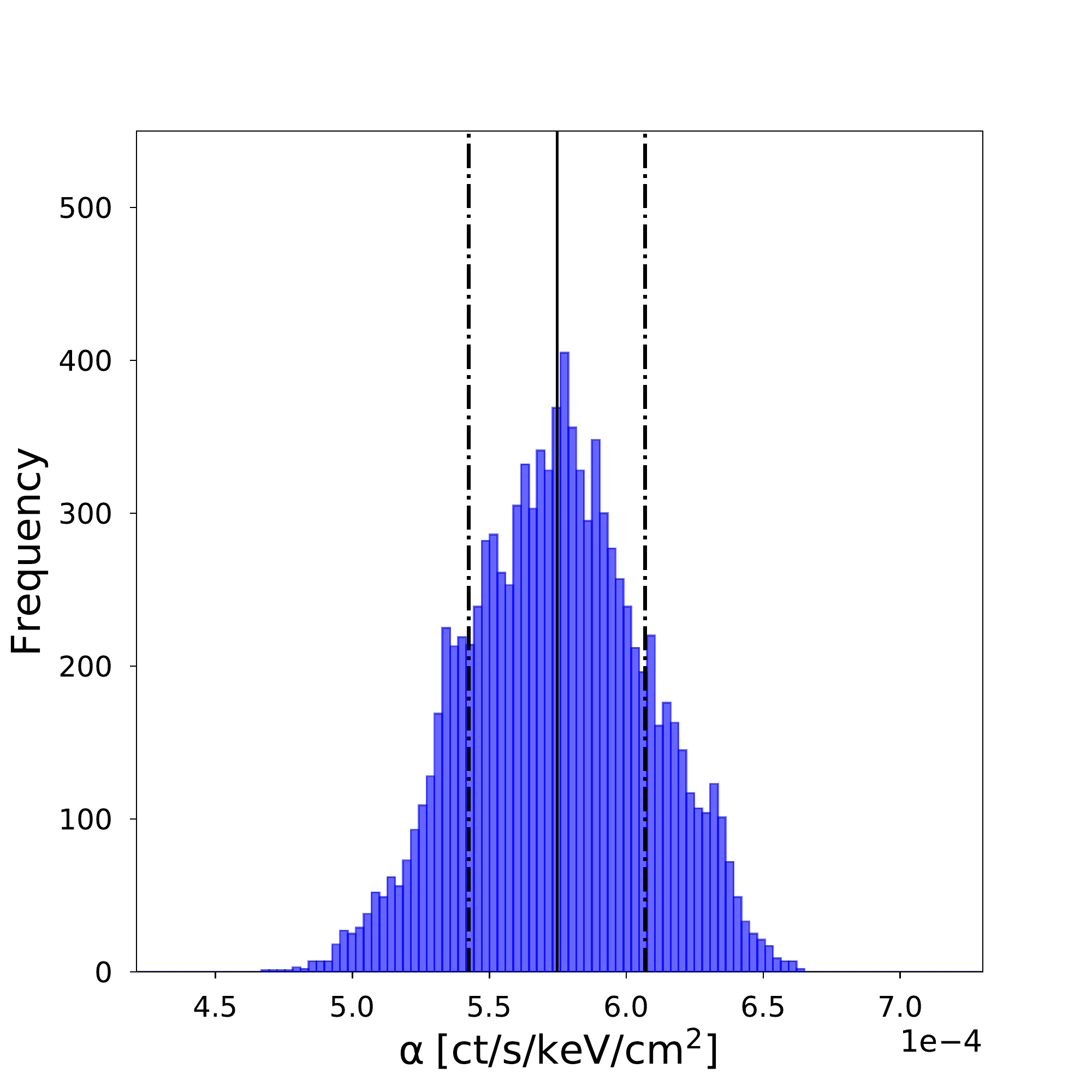}
}
\resizebox{0.8\textwidth}{!}{
\includegraphics[]{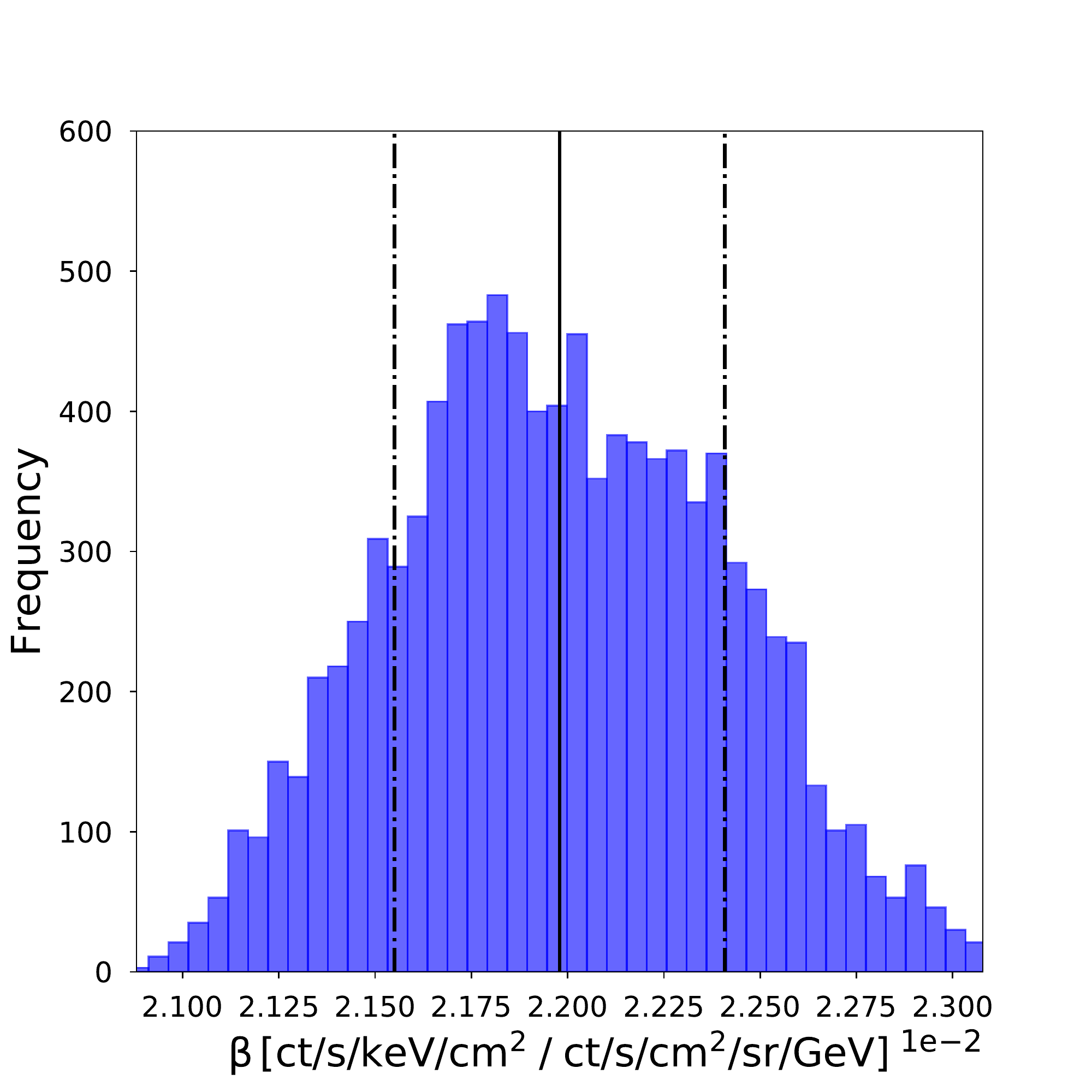}
\includegraphics[]{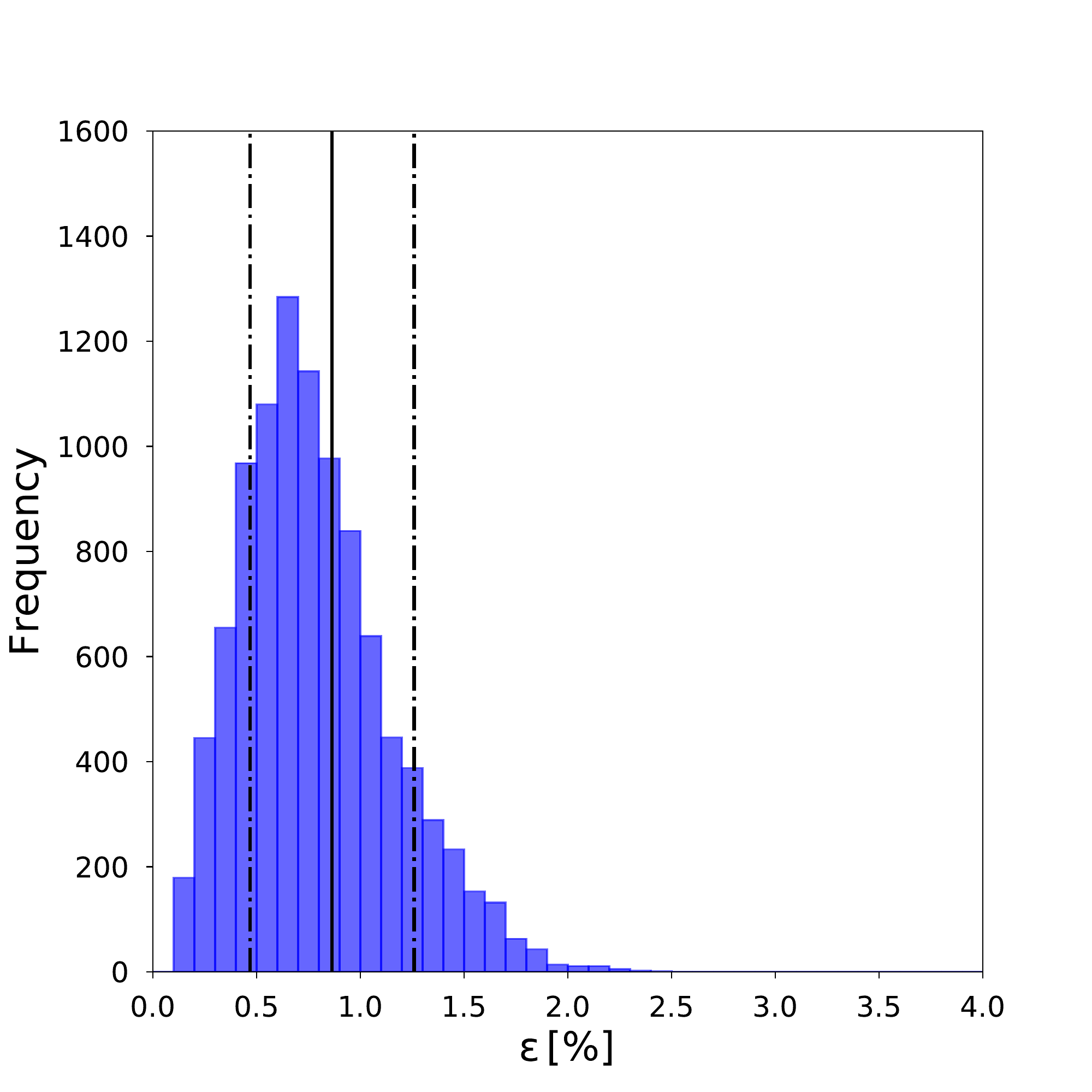}
}
\end{center}
\caption{ Correlation of the \xmm\ MOS2 \emph{outFOV} rate with the SOHO EPHIN proton flux at 1040 MeV. In the top left panel the data and the best fit relation are shown, together with the shaded area highlighting the intrinsic and total scatter. In the other panel the samples of the posterior distribution of the three fitted parameters as provided by the $linmix$ code are shown: the intercept $\alpha$ (top right panel), the slope $\beta$ (bottom left panel) and the percentage intrinsic scatter $\epsilon[\%]$(bottom right panel).}
\label{fig:9}%
\end{figure*}

\begin{figure*}[hb]
\begin{center}
\resizebox{0.8\textwidth}{!}{
\includegraphics[]{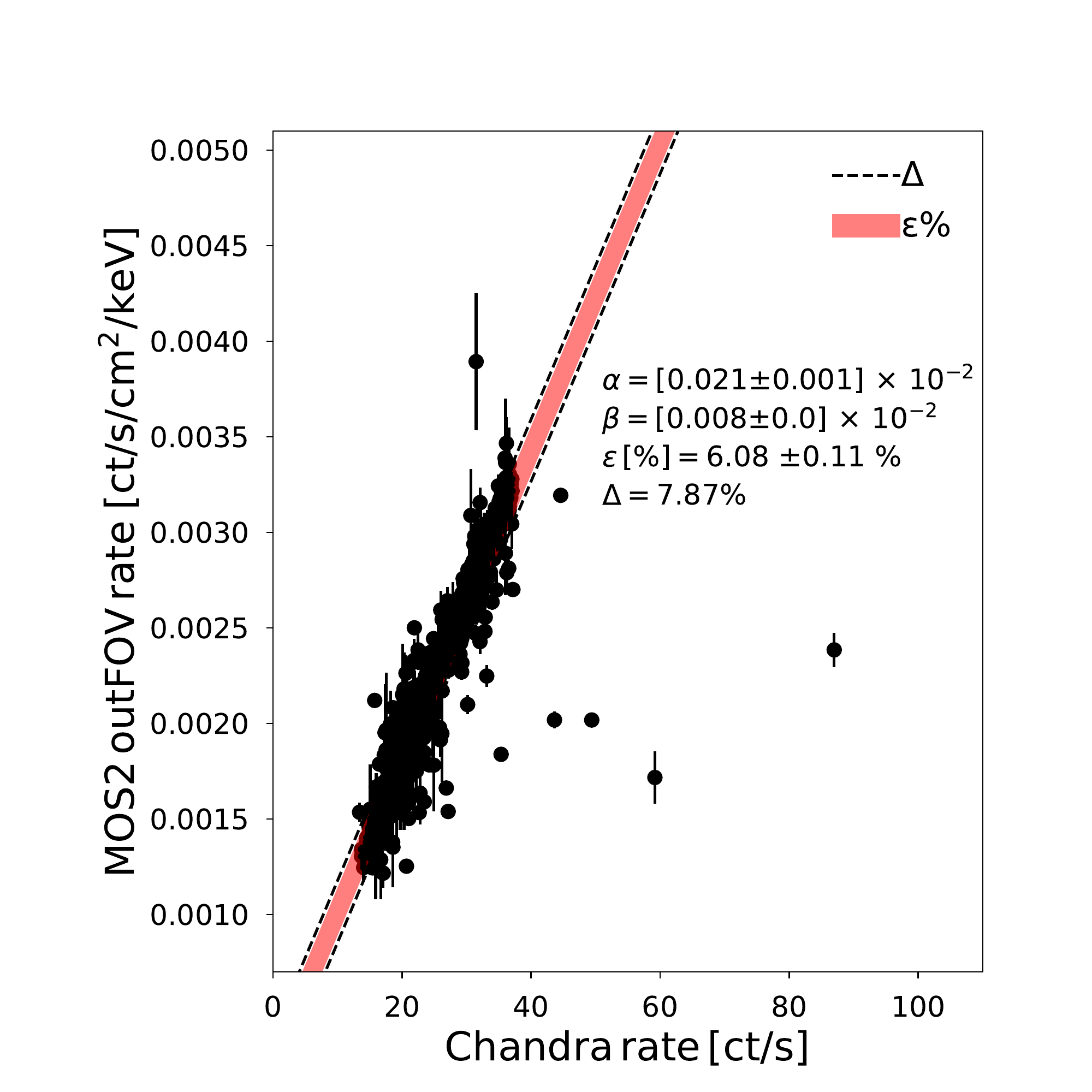}
\includegraphics[]{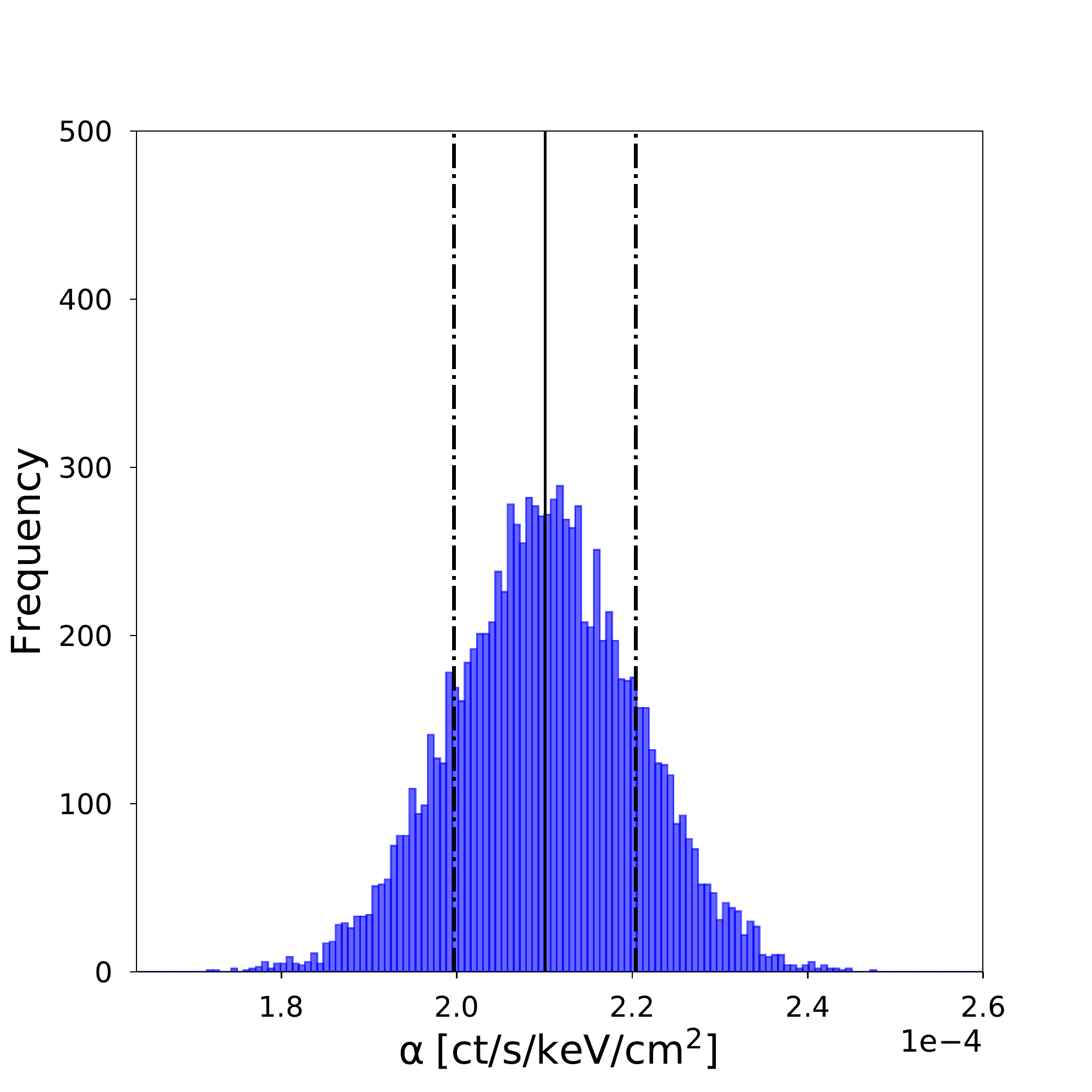}
}
\resizebox{0.8\textwidth}{!}{
\includegraphics[]{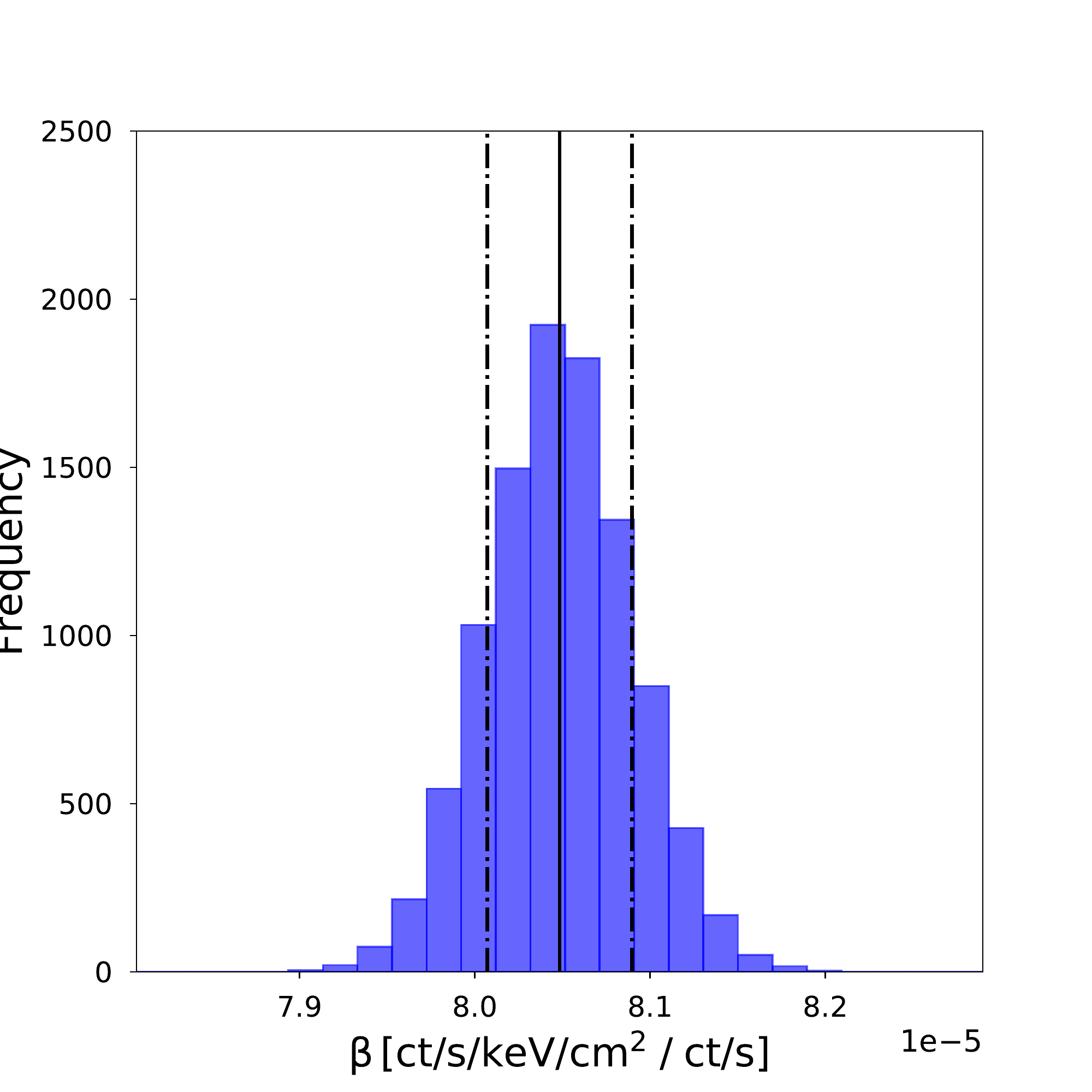}
\includegraphics[]{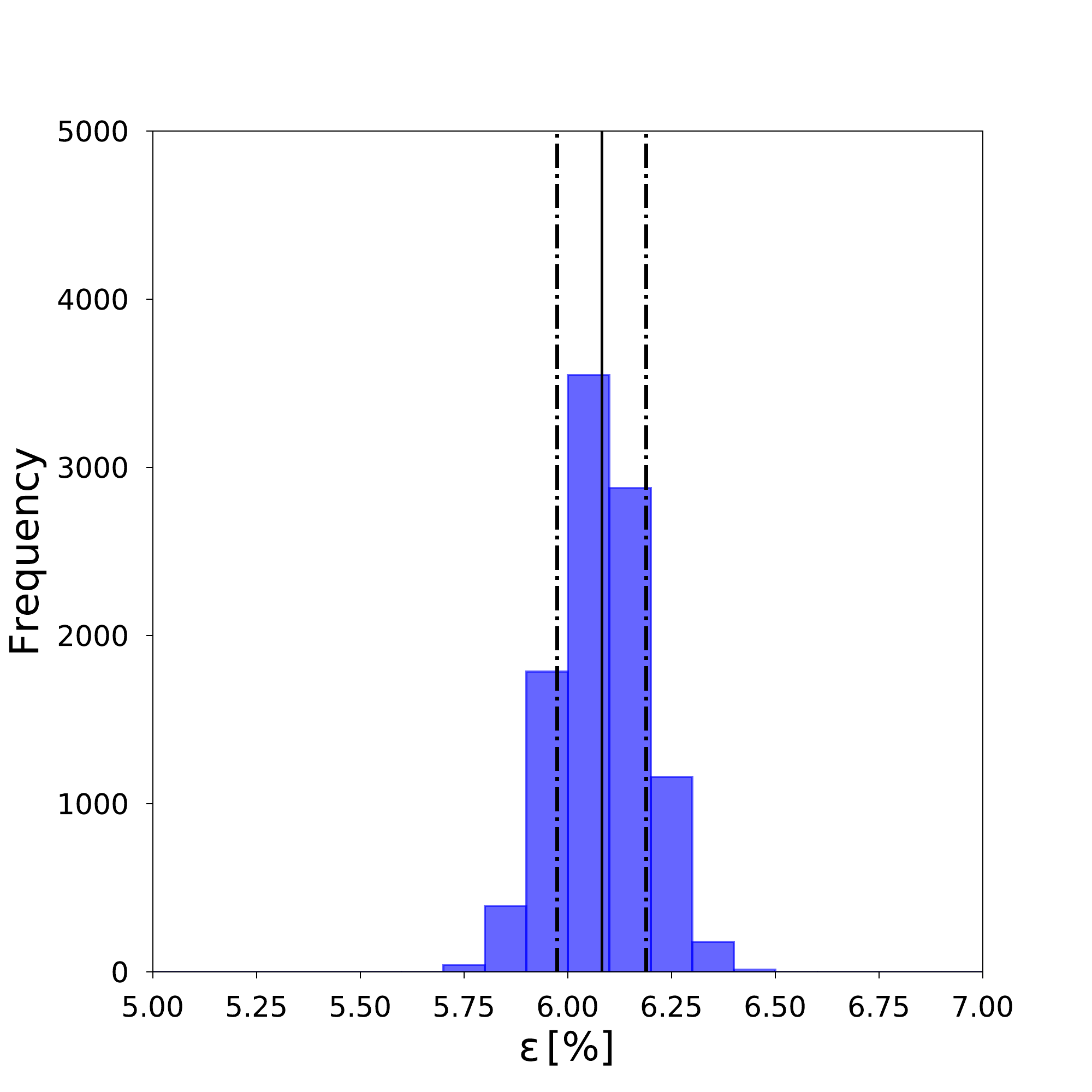}
}
\end{center}
\caption{Correlation of the MOS2 \emph{outFOV} rate with the Chandra ACIS-S3 reject rate. The outline of the figure is the same as the one of Figure \ref{fig:9} with the data, the best fit relation and the total and intrinsic scatter shown in the top left panel and the posterior for the three fitted parameters of the relation in the other panels: the intercept $\alpha$ (top right panel), the slope $\beta$ (bottom left panel) and the percentage intrinsic scatter $\epsilon[\%]$(bottom right panel).}
\label{fig:10}
\end{figure*}

\subsection{EPIC Radiation Monitor}
\label{sec:RADMONdata}
There is an instrument on board \xmm\ that registers the total count rate and basic spectral information of the  background generating radiation, it is the EPIC Radiation Monitor (ERM) \citep{Boer.ea:95}.
Its main objective is to issue a warning when the intensity of the radiation exceeds a certain level in order to trigger the shutdown of the scientific instruments and protect them from high levels of radiation.  
The ERM consists of two detectors, the low energy proton and electron unit (LE) and the high energy particle 
unit (HE). All the units are based on Silicon diodes, which record the energy loss in the material. 
The HE consists of two Si junctions equipped with electronics which registers single events (a particle passing through one junction and not the other) and the result is the counting of pulses for a selected energy band, the High Energy Single (HES) rate.
Spectral information is provided by registering the coincidence events for particles which deposit energy in both junctions (High Energy Coincidence, HEC, rate).
The ERM is operated in two different modes, fast and slow: for the slow mode a complete spectrum is
provided every 512 s whereas in the fast mode the spectrum is provided every 4 s. Counting single rates
are provided every 4s in both modes and these rates have been used in this work.

We made use of the HES0 rate which represents the number of counts accumulated over the all 256 channels of the instrument \citep[as done in a previous study,][]{Gastaldello.ea:17}. The HES0 is sensitive to protons in the 8-40 MeV range and electrons in the 1-1.75 MeV range. 
For an additional description of the ERM see this link\footnote{http://www.cosmos.esa.int/web/xmm-newton/radmon-details}.
The data are available at this link\footnote{https://www.cosmos.esa.int/web/xmm-newton/list-of-tc-radmon}. 
An example of an ERM light curve is shown in Fig.\ref{fig:7} in the various single mode events 
together with the altitude reached during the orbit by the satellite. This plot highlights the key features exploited in this work. The high ERM rates in the HES0 channel are driven by the passage in the outer electron belt \citep{Metrailler.ea:19}. The outer boundary of the outer electron belt  varies in time for two reasons. The first physical reason is that occasionally it can grow rapidly when solar eruptions reach the Earth magnetosphere and then deflate \citep[see the review by][]{Baker.ea:18}; the second and more relevant for our purposes is due to the fact that the \xmm\ orbit goes through phases of
higher and lower ellipticities with different
altitudes of apogee and perigee \citep[see e.g. Fig. 1 of][]{Walsh.ea:14}.
In every revolution an observing window is defined during which the radiation level is low enough to use
the instruments. \xmm\ turns off the scientific instruments before entering the belts and turns
them back on after exiting the radiation belts.
The ERM rates for the rest (and great majority of the time) of the orbit reflect the intensities of the Galactic Cosmic Rays. 
We therefore used the median of the ERM HES0
count rate in each \xmm\ orbit (and the median absolute deviation for its error) as a proxy for the background rate encountered by the MOS2 instrument (and in general for all the \xmm\ scientific instruments) during their operations. The plot of this quantity is shown in Fig.\ref{fig:8} alongside with the monthly number of sun spots which is a useful and easy proxy for the solar activity\footnote{data taken from http://sidc.oma.be/silso/datafiles}. The plot shows a clear anti-correlation of the ERM rate with 
the solar activity because the latter modulates GCRs. It also highlights the effectiveness of the median in removing the passage in the belts but not periods of enhanced count rates 
associated to SEP events.
\section{Correlations of the MOS2 \emph{outFOV} rate with external data-sets}
\label{sec:corrsolarcycle}

The trend with time of the MOS2 \emph{outFOV} rate (see Fig.\ref{fig:4}) shows a remarkable  similarity with the ones of the EPHIN data 
(see Fig.\ref{fig:5}), the Chandra data (see Fig.\ref{fig:6}) and the ERM data (see Fig.\ref{fig:8}). To obtain a quantitative comparison
we performed a Bayesian linear regression \citep{Kelly:07} to estimate a linear relation between the MOS2 
\emph{outFOV} rate and the EPHIN flux (in Section \ref{sec:xmmephin}), between the MOS2 \emph{outFOV} rate and the 
Chandra ACIS-S3 rejected high energy rate (in Section \ref{sec:xmmchandra}) and between the MOS2 \emph{outFOV} rate and
the ERM rate in the HES0 channel (in Section \ref{sec:xmmerm}). The Bayesian approach has the advantage of treating the intrinsic scatter as a free parameter and it accounts for measurement errors in both the dependent and independent variables.
We fitted relations of the form:

\begin{equation}
    \hat Y = \alpha + \beta\, \hat X  + \epsilon ,
    \label{eqn:correl}
\end{equation}

where $X= \hat X + \sigma_{X}$ and $Y= \hat Y + \sigma_{Y}$ with $X$ and $Y$ the measurements and $\sigma_X$ and $\sigma_Y$ their respective measurement errors. The free parameters are the intercept, $\alpha$, the slope, $\beta$ and the intrinsic scatter about the relation, $\epsilon$. We used the publicly available IDL version, $linmix\_err.pro$, of \cite{Kelly:07} code which provides samples from the posterior distribution of the three free parameters. We will quote as best fit 
values and their errors for the parameters of the regression fit the mean and standard deviations of the posterior samples provided by the
code. For the intrinsic scatter we will quote a percentage scatter, $\epsilon[\%]$, referring to $\epsilon/\,med(Y)$, the estimated scatter divided by median of the Y values. 
As the estimate of the intrinsic scatter can be biased in presence of errors which are overestimated, we will also quote the root mean square deviation (RMSD), $\sqrt{\sum_{1}^{n} (\hat Y -Y)^2/n}$ where $n$ is the number of data points, in percentage terms, $\Delta [\%] = \rm{RMSD}/med(Y)$. This is a measurement of the total scatter thus providing an upper limit to the value of the intrinsic scatter.

To convert from the units seen in the previous plots for the MOS2 \emph{outFOV} rate (re-normalized to the area of the FOV) of cts s$^{-1}$ to the units used in the next section and plots of cts s$^{-1}$ cm$^{-2}$ keV$^{-1}$ we used an energy range of 3.7 keV and a geometrical area of $29.078$ cm$^2$ for the MOS2 FOV \citep{Marelli.ea:17}.
\subsection{The correlation of the MOS2 \emph{outFOV} with SOHO EPHIN data}
\label{sec:xmmephin}
The GCR-induced background has been measured through the analysis of the corners of the MOS2 detector (MOS \emph{outFOV}) not exposed to the sky where neither X-ray photons nor soft protons penetrate. 
We matched the MOS \emph{outFOV} data with the monthly-averaged EPHIN fluxes by estimating the monthly \emph{outFOV} count-rate obtained summing all the counts accumulated during that month and dividing it by the corresponding summed exposure time.
The correlation with EPHIN fluxes for protons with energies greater than 681 MeV is very good and we show as an example the correlation with the energy of 1.04 GeV (see Figure \ref{fig:9}). The relation features a very low intrinsic scatter of about $0.8\pm0.4$\%. Being aware of the possible underestimate of this quantity by the conservative estimate of the EPHIN systematic error, we also report the total scatter which is 2.6\%, still very low.

\subsection{The correlation of the MOS2 \emph{outFOV} rate with Chandra ACIS-S3 reject rate}
\label{sec:xmmchandra}
The same qualitative time behavior is also seen in the Chandra ACIS-S3 reject rate, a proxy for the unfocused background component in Chandra (see Figure \ref{fig:6}). 
We cross-correlated the two datasets by matching the time of the stowed Chandra observation with the corresponding \xmm\ revolution which is the time bin used in Figure \ref{fig:10}. 
Notwithstanding some outliers still present even after the removal of SEP events, the correlation is very good also in this case: the linear relationship connecting the two rates has an intrinsic scatter of $5.3\pm0.1$\% and a total scatter of 7\% (see Figure \ref{fig:10}).

\begin{figure*}[htb]
\includegraphics[width=0.49\textwidth]{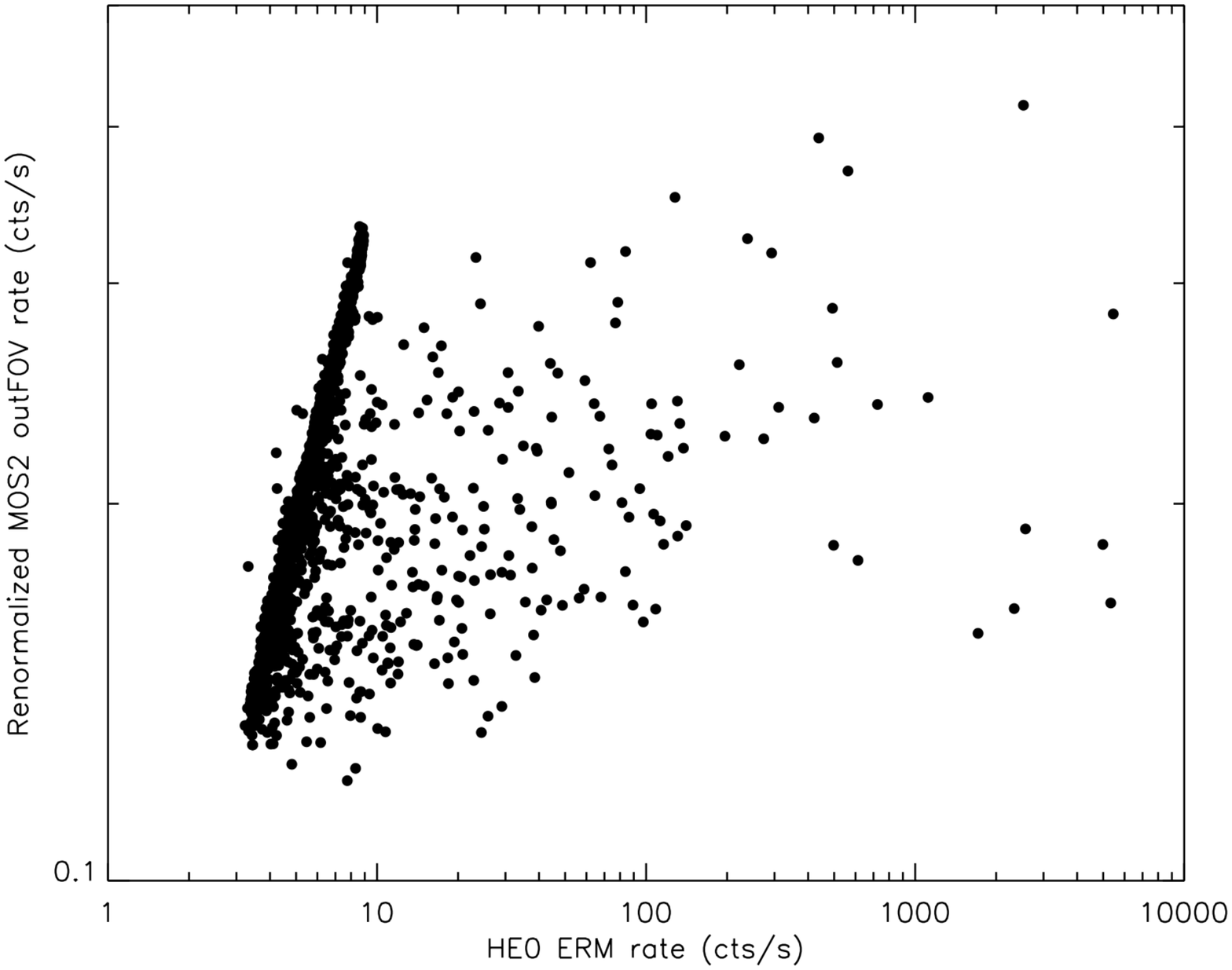}
\includegraphics[width=0.49\textwidth]{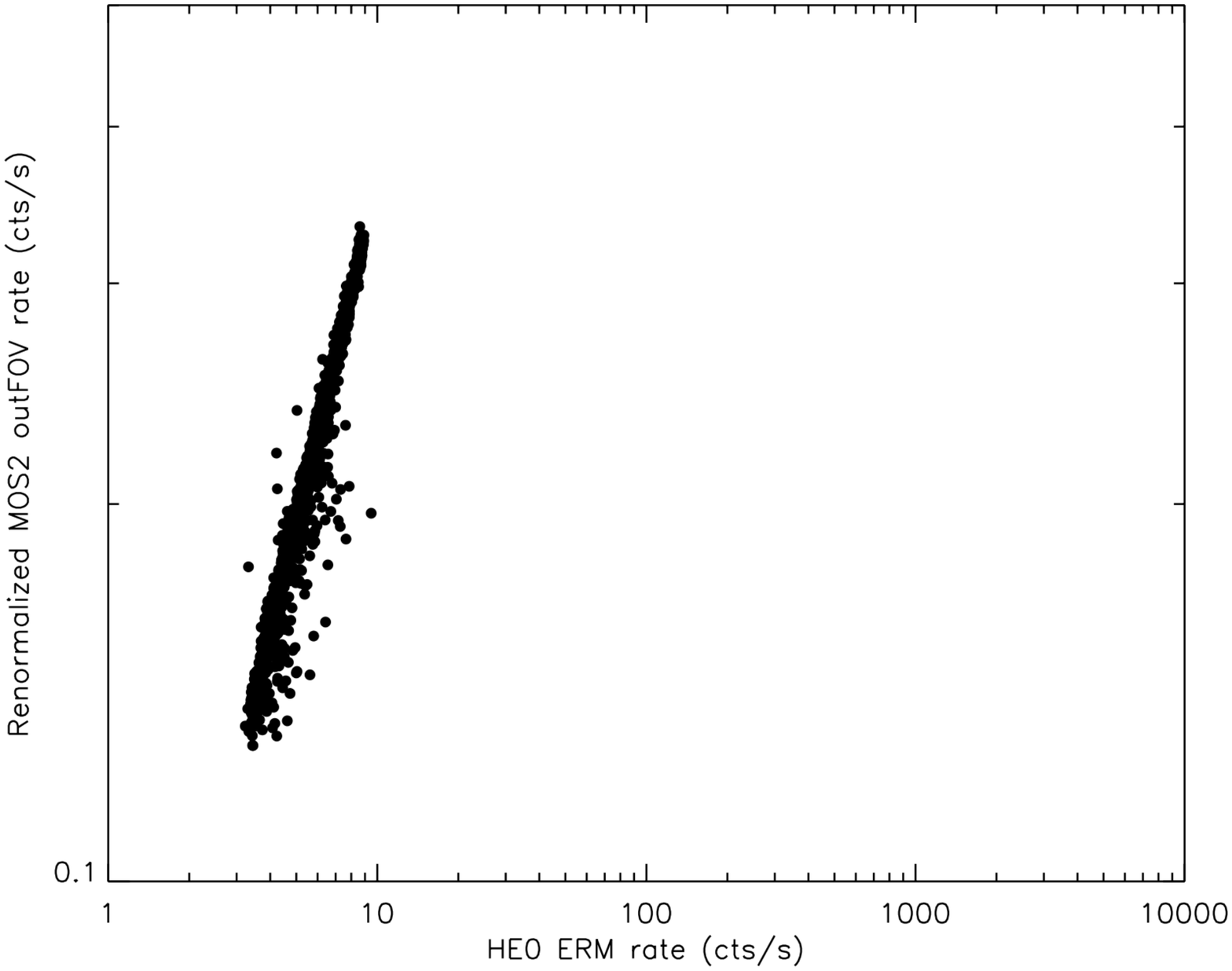}
\caption{Left Panel: Comparison of the average MOS2 \emph{outFOV} rate with the median ERM HES0 rate in the same \xmm\ orbit.
Right Panel: Same as the left panel but omitting revolutions flagged as affected by SEP events.}
\label{fig:11}
\end{figure*}

\subsection{The correlation of the MOS2 \emph{outFOV} rate with ERM}
\label{sec:xmmerm}

The plot shown in Fig.\ref{fig:7} highlights the fact that taking the median of the count rates in the ERM HES0 channel during the orbit is effective in removing features due to passage in the belts.
This is not the case for periods of enhanced count rates lasting for more than an 
\xmm\ orbit and associated to SEP events, see Fig.\ref{fig:2} for an example.
Belt passages and SEPs are periods where the count rates in the HES0 
channel do not reflect the GCR particle population.
This is highlighted by the plot in Fig.\ref{fig:11} (left panel) where we simply compare the orbit-averaged MOS2 \emph{outFOV} rate shown in Fig.\ref{fig:1}
with the median of the HES0 count-rate in the same orbit. The HES0 ERM count-rate varies by more than two orders of magnitude however the unfocused background varies at most by a factor of two. 
If we exclude the \xmm\ orbits flagged as affected by SEP events as discussed in section \ref{sec:XMMdataset}
we obtain a cleaner correlation as shown in the right panel of Fig.\ref{fig:11}. The remaining outliers are quite likely associated to SEP events as the time duration of a SEP event can be underestimated in some cases by satellites on low Earth orbit \citep[see the example shown by][]{Gastaldello.ea:17}. The same explanation can be put forward for the outliers in the correlation between the MOS2 \emph{outFOV} rate and the Chandra ACIS-S3 reject rate (see Section \ref{sec:xmmchandra}).

\begin{figure*}[hb]
\centering
\begin{center}
\resizebox{0.8\textwidth}{!}{
\includegraphics[]{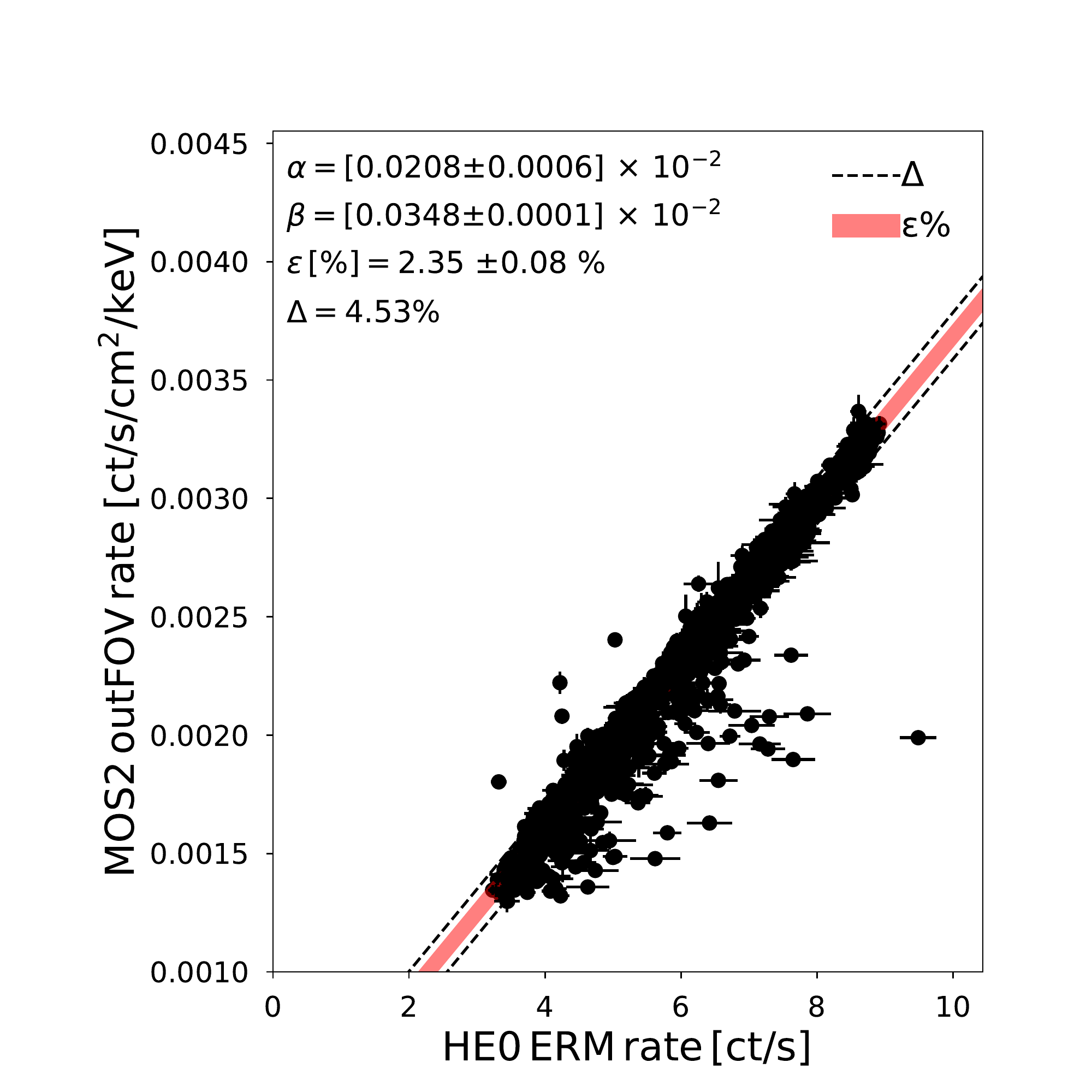}
\includegraphics[]{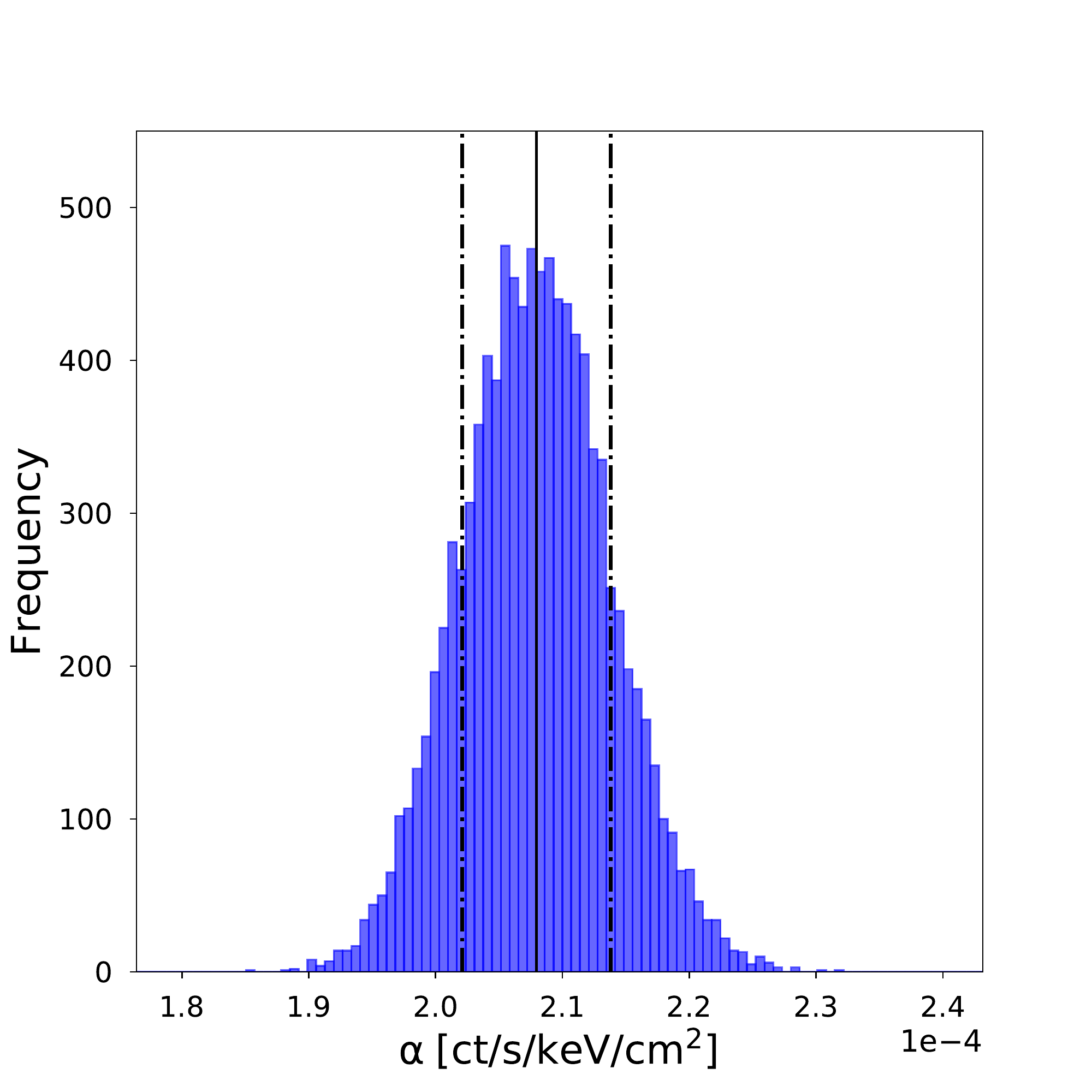}
}
\resizebox{0.8\textwidth}{!}{
\includegraphics[]{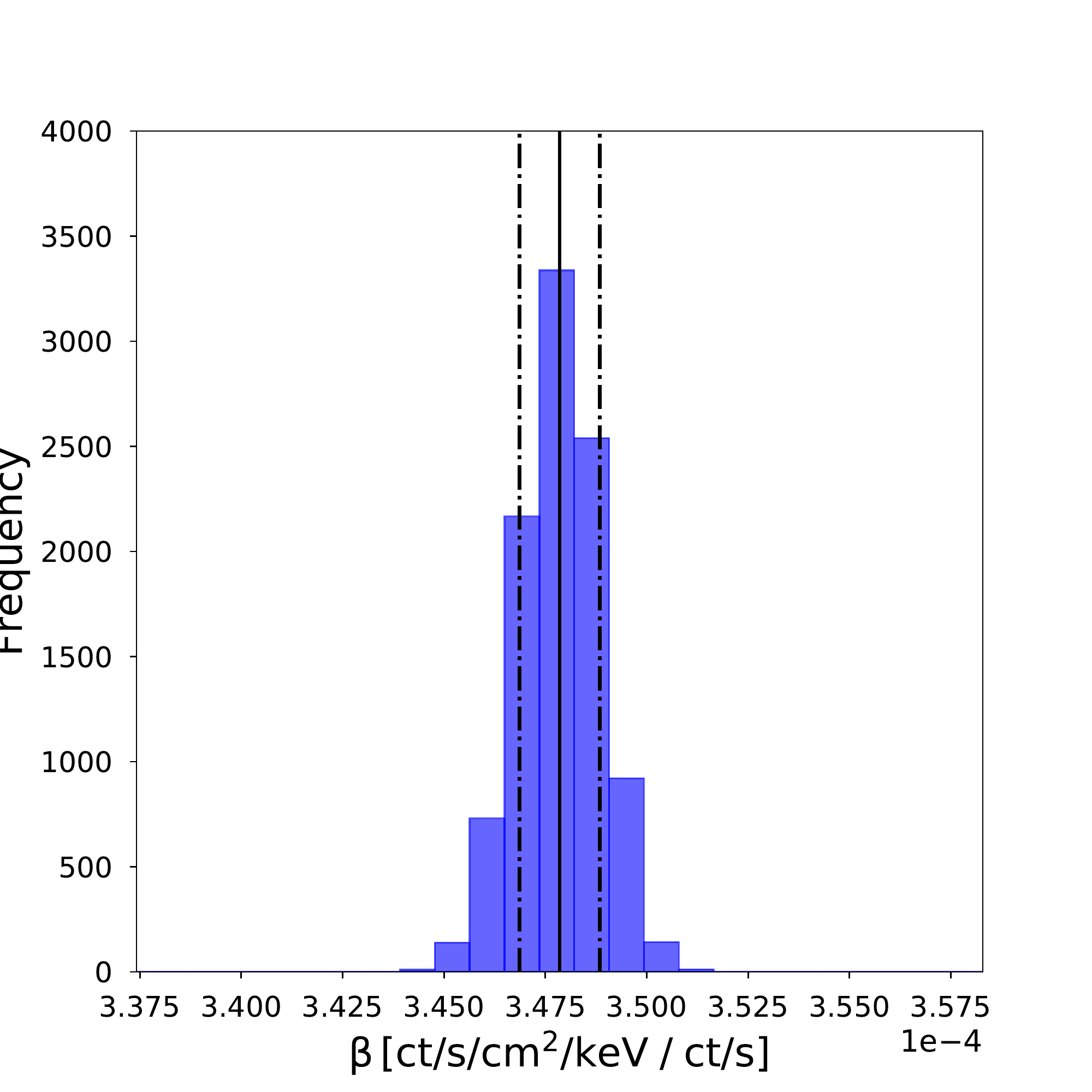}
\includegraphics[]{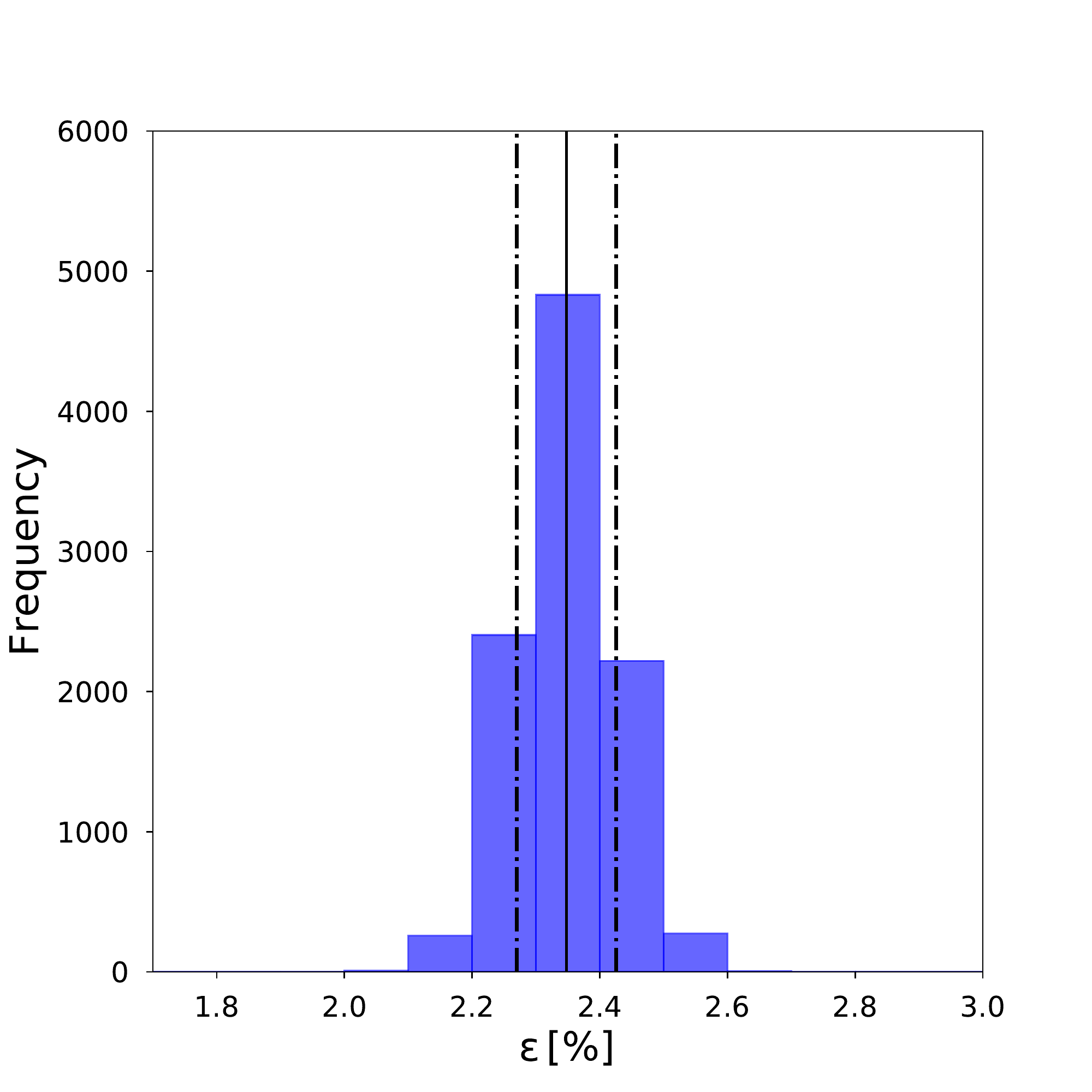}
}
\end{center}
\caption{Correlation of the MOS2 \emph{outFOV} rate with the ERM HES0 rate. The panels show the quantities in a similar way as for Figure \ref{fig:9} and figure \ref{fig:10}.}
\label{fig:12}
\end{figure*}

With this filtering the scatter in this correlation becomes low as quantified by the linear fit shown in Fig.\ref{fig:12}: the intrinsic scatter is at the level of $2.3\pm0.1$\% and the total scatter is the level of 4.5\%.


\begin{figure*}[htbp]
\includegraphics[width=0.45\textwidth]{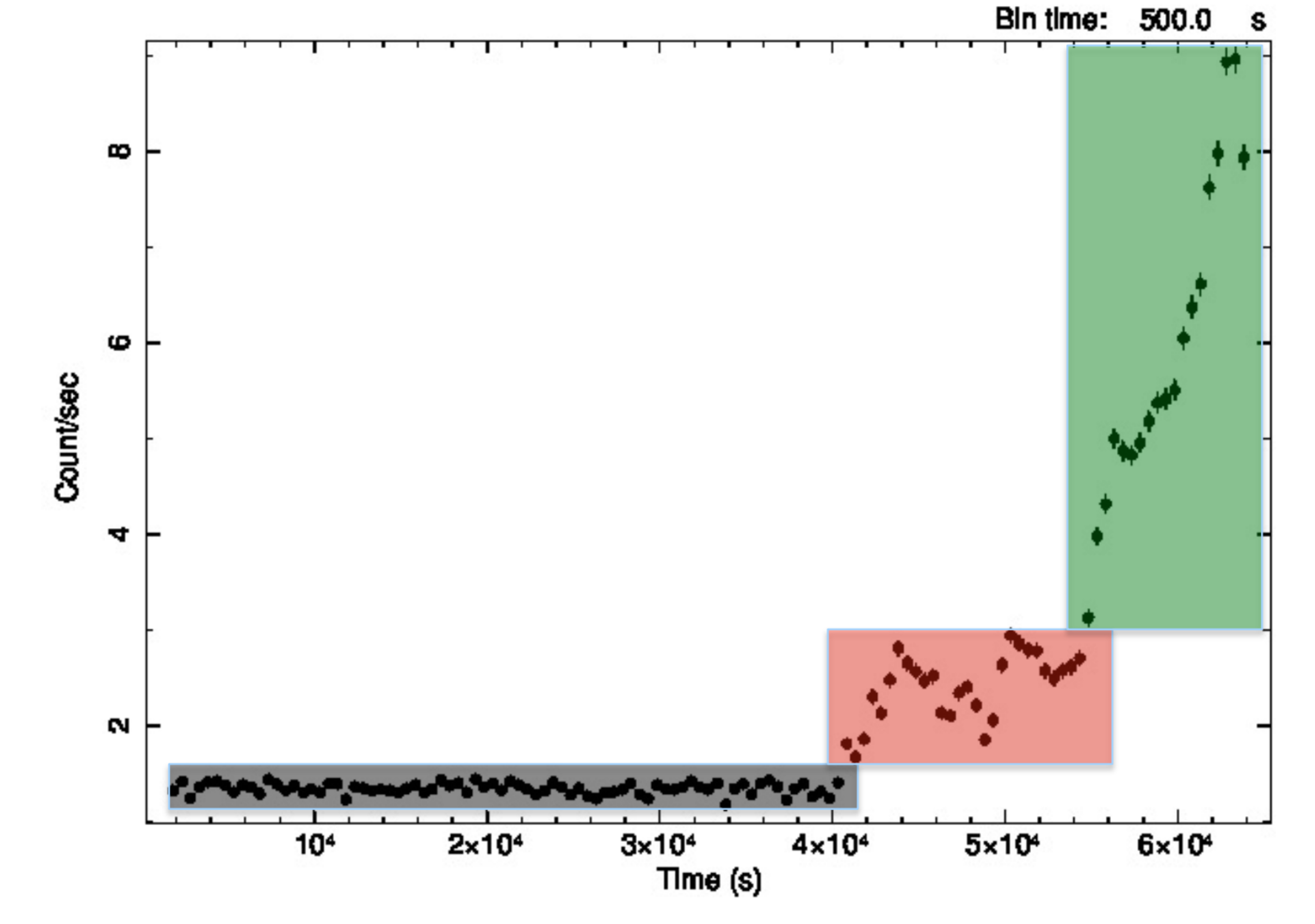}
\includegraphics[width=0.45\textwidth]{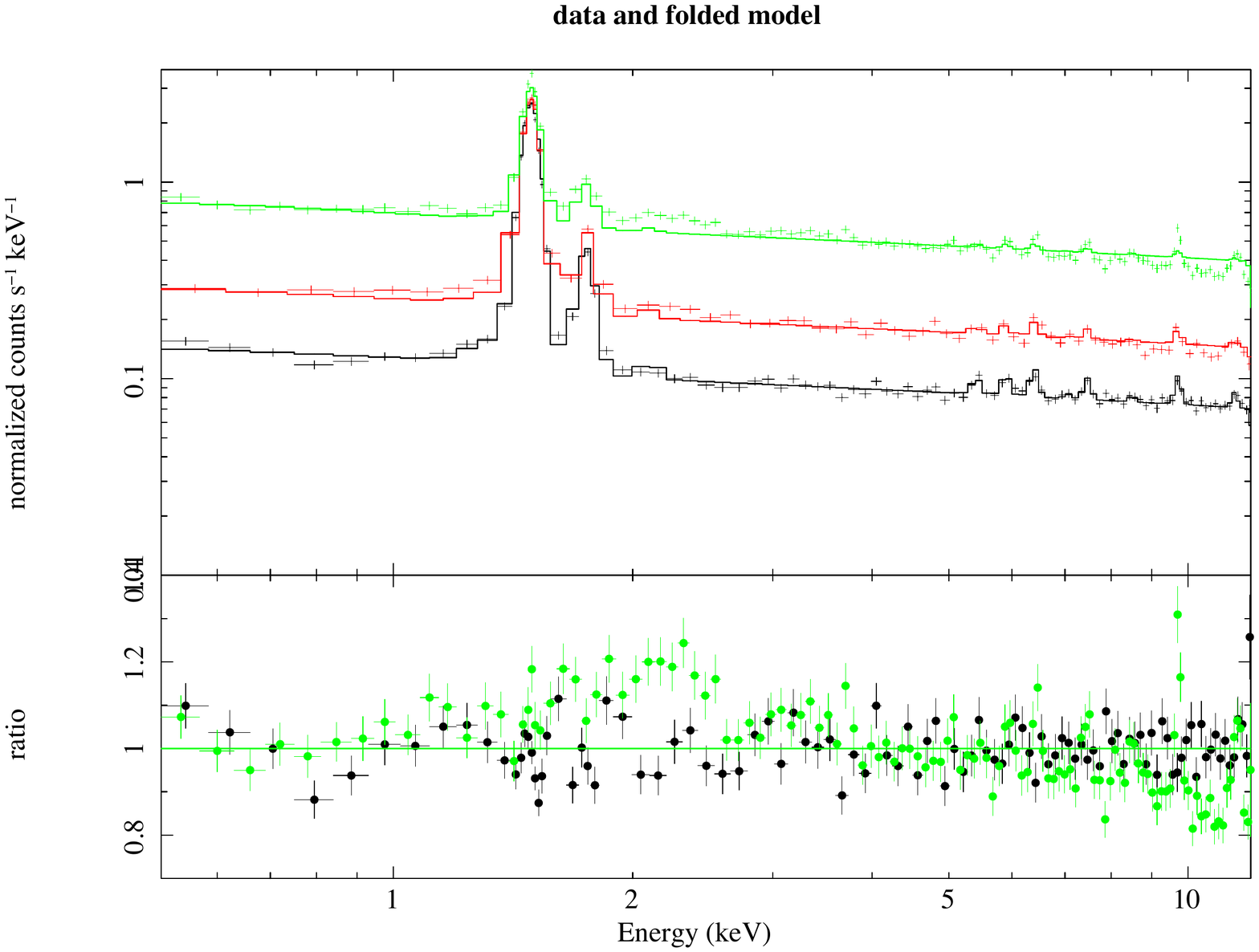}
\caption{Left Panel: The count rate of the entire MOS2 detector in the energy band 0.5-12 keV as a function of time during the closed observation with obsid 0510780101 S005. The three time intervals are highlighted for which spectra have been extracted (see right panel)
Right panel: Spectra extracted from the different time intervals shown in the left panel with the same color coding: low count-rate (black), intermediate count-rate (red) and high count-rate (green). The best fit broken power-law plus Gaussian fluorescent lines of the low-count-rate spectrum is applied to the three spectra to highlight the change in the slope in the continuum, as also shown by the residuals of the model of the low and high count-rate spectra.}
\label{fig:13}
\end{figure*}


\begin{figure}[htbp]
\includegraphics[width=1.0\textwidth]{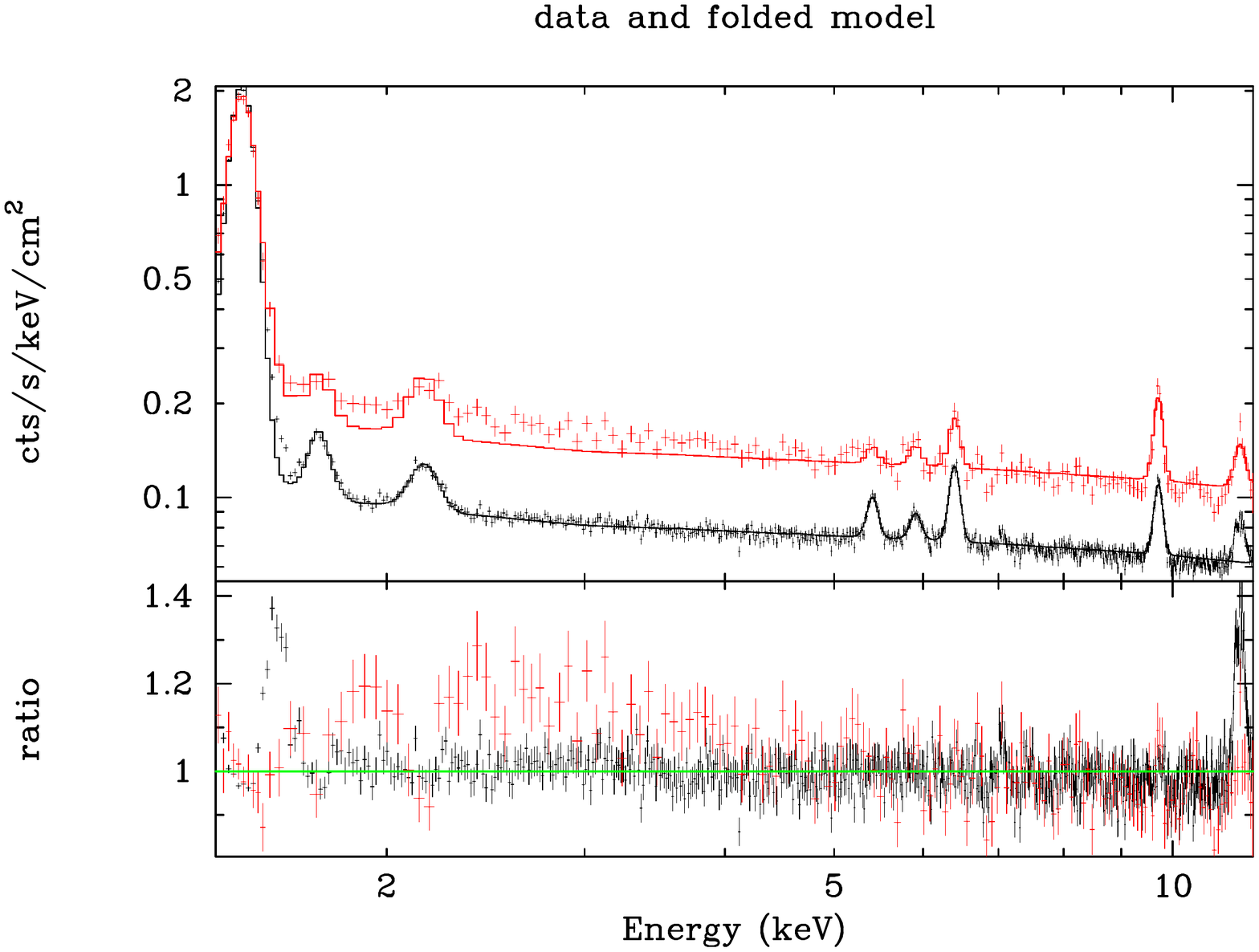}
\caption{MOS spectra accumulated from the entire revolution, in black, and from phase $>$ 0.5 and times bins with rate larger than 1.6 times the median revolution rate, in red. 
The canonical background model is fit to the first spectrum and then applied to the second. 
In the bottom panel, where we show residuals in the form of a ratio of data to model, we can observe how the spectrum collected during intensity enhancements at the end of the revolution is steeper than the typical background spectrum. }
\label{fig:15}%
\end{figure}

\section{Enhancements in the last phases of the \xmm\ orbit}
\label{sec:electronbelts}

Having established that  enhancements observed in the MOS2 \emph{outFOV} can, in most instances, be attributed to SEP events, we are left with the question of the origin of  those occurring at the end of \xmm\ orbits.
To address this point we have performed a case study of the closed observation taken during the routine calibration observation in revolution 1413 (obsid 0510780101).
Closed observations at the end of the orbit can last more than the science exposures and data
can be extracted  by the entire field of view as they are not contaminated by soft protons. The light-curve 
(see the left panel of Fig.\ref{fig:13}) shows the enhancement and has been divided in three time intervals: a  low intensity behavior range seen throughout the typical science observations, a medium intensity range and a high intensity range. Spectra taken at these different levels of intensity are shown in the  right panel of Fig.\ref{fig:13} with the same color coding. The canonical empirical model used to fit the GCR-induced background usually present in the \xmm\ data (broken power law model plus Gaussians for the fluorescence lines) is applied to the spectra extracted during the three time intervals. A statistically significant change in the high energy slope of the continuum is evident, going from the 
typical value of the unfocused MOS background of $0.18\pm0.01$ to 
$0.35\pm0.01$ in the high state. Residuals associated to the best fit model of 
the low-intensity data are also shown, highlighting the change of slope of the
continuum.

A similar result is obtained by stacking the \emph{outFOV} data from the last few (of the order of 5-10) ks of data at the end of the each orbit. If we accumulate spectra for time bins where the count rate exceeds at least by a factor of 1.6 the median count rate of the revolution (as a convenient operational way to identify the excess at the end of the orbits) we obtain the spectrum shown in red in Fig.\ref{fig:15} which bears strong similarities with the one plotted in green in Fig.\ref{fig:13}: the high energy slope is $0.39 \pm 0.02$ in good agreement with the measurement performed in the high state of the case study of the closed observation. 
A discussion on the implications of these results will be provided in Sect. \ref{sec:sub_elec}.


\section{Discussion}
\label{sec:discussion}

Analysis of the temporal variability of the MOS2 
\emph{outFOV} data and of ancillary data sets has allowed us to uncover significant correlations. In this section we will discuss the implications of our findings. 

\subsection{Correlation with GCR protons}

The tight correlation between the un-concentrated MOS2 background rate and the EPHIN proton flux, see Fig.\ref{fig:9}, strongly suggests the latter is responsible for the former.
This is consistent with findings from detailed Geant4 simulations of the interaction between high energy protons and the ATHENA WFI experiment \citep{Kienlin.ea18}, which show  the bulk of unrejected background is indeed produced by CR protons.
Moreover, the strong correlation observed between EPIC MOS2 and Chandra ACIS S3 background rates, see Fig.\ref{fig:10}, can be understood in terms of the common dependency of both on high  energy proton fluxes.
 
Since SOHO EPHIN is in orbit around L1 and \xmm\ and Chandra are on High Earth Orbits, the variation in Cosmic Ray protons fluxes between these two locations must be very modest. This is confirmed by the good correlation  found when comparing proton fluxes from  EPHIN SOHO with those recorded from the radiation monitor on board Chandra \citep[essentially a copy of the EPHIN; see][]{Kuehl.ea:16}. These results are in agreement with our current understanding of how Cosmic Rays vary within the Heliosphere, \citep[e.g.,][]{Potgieter.ea:17}. 
These findings complement and extend those presented in a recent paper \citep{Marelli.ea:21}  where we showed a  correlation  between SOHO EPHIN proton fluxes and EPIC pn background rates.

This fact is further strengthened by the similar tight correlation found between the \xmm\ MOS 2 background and the Chandra background (See Sect.\ref{sec:xmmchandra}) which points to the common origin for the background related to GCR protons for CCDs which are in similar orbits.

  Having established the MOS2 background is mostly driven by high energy protons, we turn to the looser correlation found between ERM rates and the MOS2 background rate (see Fig.\ref{fig:11} left panel).
The ERM sensitivity peaks around a few MeV, at these energies protons of solar origin provide a significant and highly variable contribution. Removal of the most obvious SEP can improve 
the correlation (see Fig.\ref{fig:11} right panel) at the cost of losing the ability to monitor the particles during SEP events. Moreover such a strategy would likely require the use of an external experiment to signal time intervals affected by SEPs. These considerations show how traditional radiation monitors, such as the ERM and the ESA Standard Radiation Environment Monitor \citep[SREM, e.g.][]{Evans.ea:08} are of limited value when trying to characterize the high energy particle component responsible for the instrumental background of X-ray detectors. This can also be understood because the goal of on-board radiation monitors is the protection of the scientific payload instruments in case of excessive radiation. Clearly, instrumentation sensitive to higher energy protons, such as the SOHO EPHIN, is far more suitable.
These findings have lead us to propose an ATHENA High Energy Particle Monitor (AHEPaM) operating at the high energies populated by the particles responsible for the un-concentrated ATHENA WFI and XIFU background.

\subsection{Electron Contribution}
\label{sec:sub_elec}
In Sect.\ref{sec:electronbelts} we have shown that at the end of revolution 1413 the background rate registered in the MOS2 corners suffered a substantial increase (Fig.\ref{fig:13}, left panel) accompanied by a change in spectral shape (Fig.\ref{fig:13}, right panel).
Since the increase occurred at a time when \xmm\ was entering the outer electron belts \citep{Metrailler.ea:19}, we attribute the spectral variation to a change in dominant contribution to background generating particles from cosmic ray protons to outer belt electrons. 
Beyond this single case evidence, a systematic analysis of the MOS2 corner data (see Fig.\ref{fig:15}) has shown the change in spectral shape to be typical of the last part of the orbit, when \xmm\ descends into the outer belts.
While practical use of this finding for EPIC may be limited, it provides an important cross validation of Geant4 simulations  showing 
that electrons contribute to the total WFI unrejected background \citep{Kienlin.ea18}.
Moreover, since  electrons are likely the only species for which a solar contribution to the ATHENA background might be significant, at least during very energetic SEPs,  AHEPaM will be designed to be sensitive to electrons 
with energies down to 50 MeV.

\subsection{Compton scattering of CXB hard X-ray photons}

The MOS2 vs EPHIN correlation, see Fig.\ref{fig:9}, shows evidence for a ``constant" component whose intensity is given by $\alpha$, the constant term in the linear relation of Eq.\ref{eqn:correl}. 
We interpret this as evidence for a contribution to the un-concentrated particle background that is distinct from the CR component. Similar evidence has been found when correlating the pn un-concentrated background with the EPHIN flux and is discussed in \citep{Marelli.ea:21}. 
As in that paper we estimate an upper and lower bound to the constant component by repeating the correlation analysis between MOS2 and EPHIN data using different proton channels for the EPHIN data: taking as representative the values of the correlations at 500 MeV and 1.2 GeV we find that in the former case the constant term is $\sim 1.1 \times 10^{-3}$ cts s$^{-1}$ cm$^{-2}$ keV$^{-1}$ and in the latter case the constant term is $\sim 5 \times 10^{-4}$ cts s$^{-1}$ cm$^{-2}$ keV$^{-1}$. Those values are a factor of $\sim 2.5 $ lower than the ones found for the pn.

In \citet{Marelli.ea:21} we presented a thorough discussion of possible origins of this component, the arguments presented there  apply here. Briefly, we can exclude the constant is an artifact in our data analysis because the MOS2 and EPHIN datasets have been collected and analyzed independently and cross-correlated only at the end.
We can also exclude the constant component is associated to Cosmic Rays or to activation, as it does not vary with the solar cycle.
The most likely interpretation is that the constant component is produced by 
cosmic X-ray photons with energies around 100 keV that Compton scatter electrons 
within the MOS2 detector. The up-scattered electrons provide a signal that is indistinguishable from that of an X-ray photon and, if their energy falls within the science band, contribute to the MOS2 background. The contribution is constant in time and does not vary when reorienting the satellite because it results from the integrated hard X-ray emission from the whole sky. 

\begin{figure}
\includegraphics[width=0.9\textwidth]{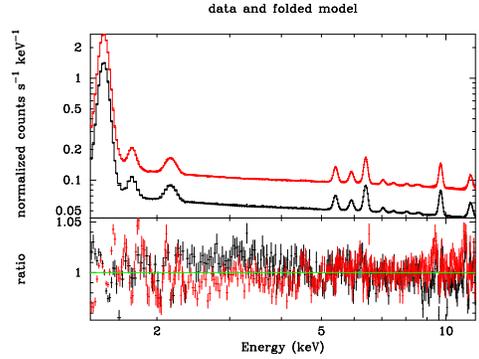}
\caption{MOS2 Spectra accumulated from the \emph{outFOV} data at solar maximum (black) and solar minimum (red). In the top panel we show a fit with a broken power-law and Gaussian lines with all the parameters linked with the exceptions of the normalizations. Computation of the Equivalent Width for the lines shows consistent values, for the stronger ones differences are below a few percent. In the bottom panel we show residuals in the form of a ratio of data over model showing overall good agreement (see discussion in the text).}
\label{fig:16}
\end{figure}

We assessed indirectly the spectral shape of the Compton component by accumulating the background from the \emph{outFOV} data at the solar maximum and at the solar minimum (see Fig.\ref{fig:16}). We fitted again the canonical GCR-induced background model of a broken power-law plus Gaussian lines with all the parameters linked between the two spectra with the exception of the normalization. The agreement is good down to a few percent, with the only significant difference found between 2 and 4.5 keV where the solar maximum spectrum shows a  $\sim$ 3\% excess with respect to the solar minimum spectrum. This can be related either to a bona-fide signature of the Compton component (emerging at solar maximum, when the contribution from GCR is minimum) or, just as likely, to some unknown systematic in the analysis. However it is relatively safe to conclude that the lack of significant changes at times when the contribution of the Compton component to the total background varies substantially suggests that it must be  similar in spectral shape to the GCR-induced one. 


Comparison of the pn vs EPHIN plot presented in \citet{Marelli.ea:21} with MOS2 vs EPHIN presented here shows a difference in the constant term.  
This can be accommodated, at least qualitatively, within the framework of the Compton model, by postulating that it relates to the difference in thickness of the active region in the two detectors, about 45 $\mu$m for MOS2 and 280 $\mu$m for pn. The ratio between the two thicknesses is about 6 which is consistent, within a factor of 2, with the ratio between the pn and MOS constant components. 
We speculate that the higher sensitivity of the MOS2 detector to the Compton component could be due to  a substantial field-free region missing in the fully depleted pn. Electrons suffering a Compton scattering in this region generate a charge cloud that could end up being at least partially collected by the CCD. Since Compton photons have energies of the order of 100 keV, even a partial charge collection could lead to an event in the science band.

An additional contribution in the MOS detector with respect to the pn comes from the time spent during readout of the signal in the storage area. The readout time equals the frame time (2.6 s) and the storage area is 3/10 of the CCD area. The storage area is shielded by a 3.2 mm thick layer of an Al alloy (Abbey and Ross, private communication) which is however basically transparent to hard X-ray photons producing the Compton signal. As a benchmark the mean free path for Compton scattering of a 100 keV photon in Al is 2.7 cm.

A more thorough investigation of the issue of the Compton induced background would require a detailed simulation of the interaction between hard X-ray photons and the EPIC pn and MOS cameras, which is beyond the scope of this paper. 



\section{Summary}
\label{sec:summary}

The main findings presented in this paper may be summarized as follows.
\begin{itemize}
    \item The tight correlation found between MOS2 background and EPHIN proton flux strongly suggests the bulk of the MOS2 un-concentrated background is generated by GCR protons.
    \item The tight correlation between the MOS2 background and the Chandra background reinforces the common origin related to GCRs for CCDs which are in similar orbits.
    \item The weaker correlation found between MOS2 background and ERM data results from the significant solar contribution to protons at the MeV energies where ERM  sensitivity peaks. Only with appropriate filtering of solar particles the correlation can be tightened.
    \item These results show how traditional radiation monitors, such as the ERM and the ESA SREM are of limited value when trying to characterize the high energy particle component responsible for the instrumental background of X-ray detectors. Instrumentation sensitive to higher energy protons, such as the SOHO EPHIN, is far more suitable.
    \item A single case study, later corroborated by a systematic analysis of the MOS2 corner data, has highlighted spectral change in EPIC MOS2 background when \xmm\ descends into the outer belts. We interpret this as evidence of a substantial increase in the electron contribution to the EPIC MOS2 background.
    \item As for the EPIC pn, for EPIC MOS we find evidence of a constant component independent of CR particles. Following  (Marelli et al. 2021) we attribute this component to Compton scattering of hard Cosmic X-ray photons in the MOS2 detector. The smaller intensity found in MOS2 with respect to pn is in qualitative agreement with the smaller thickness of the former detector with respect to the latter.
     
\end{itemize}

Findings presented in this paper and in Marelli et al. (2021) have lead to the formulation of  a conceptually new particle monitor for the ATHENA mission: the ATHENA High Energy Particle Monitor (AHEPaM) that is currently under study.

\section*{Acknowledgements}
We acknowledge support from the Horizon 2020 Programme under the AHEAD2020 project (grant agreement No. 871158). F.G., S.M., A.D.L. and A.T. acknowledge support from the INAF mainstream project ``Characterizing the background of present and future X-ray missions" 1.05.01.86.13.
This research has made use of data produced by the EXTraS project, funded by the European Union's Seventh Framework Programme under grant agreement No. 607452. The scientific results reported in this paper are based on observations obtained with XMM-Newton, an ESA science mission with instruments and contributions directly funded by ESA Member States and NASA.
The SOHO/EPHIN project is supported under Grant 50 OC 1702 by the German
Bundesministerium f{\"u}r Wirtschaft through the Deutsches Zentrum f{\"u}r Luft-
und Raumfahrt (DLR).
We would like to thank Kip Kuntz and Perry Brendan for useful discussion and advice about the treatment of the pseudo-logaritmic data compression in the ERM data. We would like to thank Tony Abbey and Duncan Ross for providing the drawings and the specifications of the MOS frame storage shield and for helpful discussions. We would like to thank Nicola La Palombara for helpful discussions and insights. 

\bibliographystyle{aasjournal}
\bibliography{gasta}

\end{document}